\documentclass[fleqn,usenatbib]{arxivpub}

\usepackage{newtxtext,newtxmath}

\usepackage[T1]{fontenc}

\DeclareRobustCommand{\VAN}[3]{#2}
\let\VANthebibliography\thebibliography
\def\thebibliography{\DeclareRobustCommand{\VAN}[3]{##3}\VANthebibliography}

\usepackage{graphicx}	
\usepackage{amsmath}	
\usepackage{siunitx}
\usepackage{subcaption}
\usepackage{color}
\newcommand{\red}{\color{black}}

\DeclareSIUnit\parsec{pc}
\DeclareSIUnit\h{\textit{h}}
\newcommand{\mpc}{\mega\parsec}

\newcommand{\defeq}{\equiv}

\newcommand{\ft}[1]{\tilde{#1}}
\newcommand{\abacussummit}{\texttt{AbacusSummit}}
\newcommand{\mean}[1]{\left\langle #1 \right\rangle}

\renewcommand{\vec}[1]{\boldsymbol{#1}}

\newcommand{\vb}{{\vec{b}}}

\newcommand{\vk}{{\vec{k}}}

\newcommand{\vq}{{\vec{q}}}
\newcommand{\vr}{{\vec{r}}}

\newcommand{\vw}{{\vec{w}}}
\newcommand{\vx}{{\vec{x}}}

\newcommand{\vz}{{\vec{z}}}

\newcommand{\vW}{{\vec{W}}}

\title[Reconstructing Initial Conditions with CNNs]{Reconstructing Cosmological Initial Conditions from Late-Time Structure with Convolutional Neural Networks}

\author[Shallue and Eisenstein]{
Christopher J. Shallue,$^{1}$\thanks{E-mail: cshallue@cfa.harvard.edu}
Daniel J. Eisenstein$^{1}$
\\
$^{1}$Center for Astrophysics | Harvard \& Smithsonian, 60 Garden St, Cambridge, MA 02138, USA\\
}

\date{Accepted 12 February 2023. Received 1 January 2023; in original form 25 July 2022}

\pubyear{2023}

\begin{document}
\label{firstpage}
\pagerange{6256--6267}
\maketitle

\begin{abstract}
We present a method to reconstruct the initial linear-regime matter density field from the late-time non-linearly evolved density field in which we channel the output of standard first-order reconstruction to a convolutional neural network (CNN). Our method shows dramatic improvement over the reconstruction of either component alone. We show why CNNs are not well-suited for reconstructing the initial density directly from the late-time density: CNNs are local models, but the relationship between initial and late-time density is not local. Our method leverages standard reconstruction as a preprocessing step, which inverts bulk gravitational flows sourced over very large scales, transforming the residual reconstruction problem from long-range to local and making it ideally suited for a CNN. We develop additional techniques to account for redshift distortions, which warp the density fields measured by galaxy surveys. Our method improves the range of scales of high-fidelity reconstruction by a factor of 2 in wavenumber above standard reconstruction, corresponding to a factor of 8 increase in the number of well-reconstructed modes. In addition, our method almost completely eliminates the anisotropy caused by redshift distortions. As galaxy surveys continue to map the Universe in increasingly greater detail, our results demonstrate the opportunity offered by CNNs to untangle the non-linear clustering at intermediate scales more accurately than ever before.
\end{abstract}

\begin{keywords}
large-scale structure of Universe -- early Universe -- methods: data analysis
\end{keywords}



\section{Introduction}

Over the past century, observational data have revolutionized our understanding of the geometry and composition of the Universe.
The present-day Universe is the result of $14$ billion years of evolution according to physical laws governing its components: dark energy, dark matter, baryonic matter, and radiation.
Although some of these components and their interactions are now well-understood, others remain relatively mysterious. In particular, the nature of dark energy and dark matter remains unknown, as do the non-gravitational interactions (if any) between dark matter and baryonic matter, photons, and other dark matter.
As we continue to gather more observational data at increasingly higher precision, we hope to learn more about the history of the Universe, the nature of its components, and the laws governing its evolution.
However, going further with observational data also requires techniques for separating out the effects of different physical processes and observational distortions on that data.

Gravity is one of the primary physical processes responsible for the rich structures we observe in the present-day Universe (e.g. galaxies and galaxy clusters). 
Gravity became the dominant force on baryonic matter at the epoch of recombination when the Universe had cooled sufficiently for electrons to combine with atomic nuclei; prior to this, interactions between photons and free electrons kept the baryons coupled to the radiation, preventing gravitational collapse through radiation pressure.
The state of matter in the present-day Universe therefore arose from an unobserved initial state at recombination, evolved forward in time primarily under the influence of gravity.
If we could remove the effect of gravity on the present-day matter distribution, revealing the initial state at recombination, we could probe physical effects that have been distorted or obscured by gravitational collapse.
A well-known example is the peak in the matter 2-point correlation function due to baryon acoustic oscillations (BAO): sound waves in the photon-baryon fluid in the early Universe that were frozen-in at recombination.
BAO measurements provide a standard ruler used by dark energy surveys to map the expansion history of the Universe and probe the evolution of dark energy \citep{weinberg_observational_2013}.
The BAO signature is measurable in the present-day matter field, but it is distorted by the evolution of galaxies away from their original locations; recovering the original matter field would allow us to measure the sound horizon with greater precision.
The initial state of matter at recombination could also probe primordial non-Gaussianity \citep{bartolo_non-gaussianity_2004} and, more broadly, might contain unanticipated discoveries about the constituents of the Universe and the laws governing them.

Unfortunately, although we can accurately model gravitational dynamics \textit{forward} in time, reversing the process to reconstruct the initial state from the final state is not straightforward.
Subregions that have collapsed into virialized systems such as galaxies have no unique inverse solution; information about the initial state has been lost. Thankfully, this has only happened at the smallest scales. Averaging over sufficiently large scales ($k \lesssim \SI{0.1}{\h\per\mpc}$), perturbations from a homogeneous density remain small and gravitational dynamics are described by first-order perturbation theory.
In this theory, the density field evolves as a linear combination of a growing mode solution, in which density increases with time, and a decaying mode solution, in which density decreases with time.
At smaller scales, however, perturbations are sufficiently large in magnitude that perturbative solutions break down.
Attempts to simulate the matter field with time reversed or numerically solve the inverse gravity problem are doomed to fail: noise and numerical errors in the final state will be amplified by the decaying mode and will eventually dominate.
One therefore needs an approach that controls or eliminates the decaying mode.

Observational constraints pose additional challenges: we cannot even measure the true matter density at the present day. By inferring distances to galaxies from their redshifts, we incur errors due to their peculiar velocities, distorting the density map in the radial direction \citep{scoccimarro_redshift-space_2004}. A particular manifestation of these redshift distortions is the finger-of-God effect, where collapsing structures appear elongated along the line of sight.
Moreover, the measured matter density is biased by our use of luminous galaxies as tracers, whereas the true matter field is comprised mostly of dark matter.

Despite these challenges, a number of \textit{reconstruction}  techniques have been developed to approximately recover the initial matter density and/or velocity fields from the late-time fields. \citet{peebles_tracing_1989,peebles_gravitational_1990} reconstructed the initial positions of galaxies in the Local Group by solving for their orbital trajectories using the principle of least action. \citet{nusser_tracing_1992} derived equations for gravitational fluid dynamics in the Zel'dovich approximation \citep{zeldovich_gravitational_1970} that could be integrated backward in time. \citet{weinberg_reconstructing_1992} proposed `Gaussianizing' the final field and then evolving it forward in time to determine the overall amplitude of the fluctuations. Others extended these approaches and proposed alternatives \citep[e.g.][]{gramann_improved_1993,croft_reconstruction_1997,narayanan_reconstruction_1998,monaco_reconstruction_1999,nusser_least_2000,goldberg_using_2000,valentine_reconstructing_2000,frisch_reconstruction_2002,brenier_reconstruction_2003}.
\citet{eisenstein_improving_2007} demonstrated that the BAO feature in galaxy surveys can be sharpened using a simple reconstruction algorithm based on the linear perturbation theory continuity equation.
We will refer to this algorithm as \textit{standard reconstruction}.
It has been extensively studied both theoretically \citep[e.g.][]{padmanabhan_reconstructing_2009,noh_reconstructing_2009} and with simulations \citep[e.g.][]{seo_improved_2007,seo_high-precision_2010,mehta_galaxy_2011,burden_efficient_2014,achitouv_improving_2015,white_reconstruction_2015,vargas-magana_clustering_2016,seo_modeling_2016}.
\citet{padmanabhan_2_2012} extended the algorithm to also correct linear-theory redshift distortions and used it to obtain a factor of 2 improvement in the error of the distance scale measured from the BAO signature.
It has become the prevailing reconstruction method for improving BAO measurements in galaxy surveys \citep[e.g.][]{anderson_clustering_2012,anderson_clustering_2014,alam_clustering_2017}.

Although standard reconstruction has proved very successful at the scales relevant to the BAO ($\sim \SI{150}{\mpc}$), there is still considerable interest in recovering the initial density field with higher fidelity at smaller scales, for example to probe non-Gaussianity \citep{mohayaee_reconstruction_2006}. Such improvements would also help BAO measurements by improving the fit of the reconstructed correlation function to a theoretical template at small scales.
\citet{schmittfull_iterative_2017} provide a detailed review and classification of published reconstruction algorithms.
More recently, iterative techniques have been developed to solve for the initial density field directly \citep{hada_iterative_2018} or to solve for the nonlinear mapping between initial Lagrangian coordinates and final Eulerian coordinates of particles \citep{shi_new_2018,wang_iterative_2020}. Meanwhile, \citet{levy_fast_2021} improved the efficiency of reconstruction when it is cast as an optimal transport problem and \citet{sarpa_bao_2019} extended the method of \citet{nusser_least_2000} for massive spectroscopic surveys.
An alternative set of approaches employ fast forward models that transform initial particle positions into final positions, either to convert reconstruction into a maximum \textit{a~posteriori} (MAP) optimization problem \citep{feng_exploring_2018} or to perform Bayesian inference by iteratively sampling initial conditions \citep{bos_bayesian_2019,kitaura_cosmic_2021}. Although these methods can in principle use arbitrary forward models, simplified approximate models are required to make the techniques computationally tractable, potentially constraining their accuracy.

The reconstruction techniques mentioned in the preceding paragraphs all derive from explicit models of gravitational dynamics.
Recent large, high-accuracy cosmological simulations make possible an alternative approach: use machine learning techniques like convolutional neural networks (CNNs) to learn the inverse transformation from final conditions to initial conditions. Mapping directly from final state to initial state sidesteps the decaying-mode issue that arises in true inverse dynamics. Moreover, one can apply observational distortions like redshift distortions and galaxy bias to the final conditions and train the model to correct those distortions simultaneously with reversing gravity.

In this work, we argue that CNNs \textit{alone} are not well-suited for reconstructing the initial density from the final density because CNNs are local models whereas gravity is a long-range force: information needed to undo local gravitational displacements is sourced over very large scales. We propose a new method that applies standard reconstruction as a preprocessing step and then uses a CNN to map the partially-reconstructed density field to the true initial density field. The preprocessing step reverses the large-scale bulk gravitational flows, transforming the remaining reconstruction task from long-range to local and making it well-suited for treatment with a CNN.
Additionally, we tackle the more difficult problem of reconstruction from a density field warped by redshift distortions -- that is, we assume that distances to objects measured along the line of sight are subject to errors due to their peculiar velocities, as would be the case in practice.
We extend our base method to effectively undo redshift distortions while simultaneously reconstructing the initial density field.
We demonstrate a significant improvement over standard reconstruction, with our method improving the range of scales of high-fidelity reconstruction by a factor of 2, corresponding to a factor of 8 increase in the number of well-reconstructed modes.

Two other recent studies have also used neural networks (NNs) for reconstruction. \citet{modi_cosmological_2018} trained a NN to generate positions and masses of dark matter halos for a given density field, which, in combination with a differentiable forward model of initial to final density, they used to find the MAP estimate of the reconstructed initial density given a set of observed halos.
Later, \citet{modi_cosmicrim_2021} trained a recurrent neural network to perform fast MAP inference of the initial density field using differentiable forward models of gravitational dynamics and halo bias.
In contrast to these studies, we train our CNN to map its input directly to initial density without explicitly modeling gravitational dynamics in the training process.

This paper is structured as follows. In Section~\ref{sec:cnns} we provide a brief overview of CNNs. In Section~\ref{sec:methods} we describe our methods, including the reconstruction algorithm we use for preprocessing, our base CNN architecture, our data set, and our training procedure. In Section~\ref{sec:results} we present the results of our reconstruction technique and describe the changes needed to extend our base method in the presence of redshift distortions. We also investigate the effects of changing the cosmology of the input data with respect to the training data. We conclude in Section~\ref{sec:conclusions}.

\section{Convolutional Neural Networks}\label{sec:cnns}

A \textit{neural network} is a type of parameterized function that transforms multidimensional data. It is comprised of a sequence of \textit{layers}, each of which performs a simple mathematical transformation on its input,  with the output becoming the input to the next layer. When many layers are chained together into a ``deep'' neural network, the function becomes extremely flexible and can represent complex transformations \citep[e.g.][]{lecun_deep_2015,goodfellow_deep_2016}.

The simplest kind of a neural network is a \textit{fully connected} neural network, in which the transformation performed by layer $i$ is
\begin{equation}
    \vx_i = \phi(\vW_i \vx_{i-1} + \vb_i),
\end{equation}
where $\vx_i$ is a vector of length $n_i$ of outputs from layer $i$, $\vx_0$ is the input data, $\vW_i$ is an $n_i \times n_{i-1}$ matrix of \textit{weight} parameters, $\vb_i$ is a length-$n_i$  vector of \textit{bias} parameters, and $\phi$ is an elementwise nonlinear \textit{activation function} (e.g. the hyperbolic tangent function $\phi(x) = \tanh(x)$)
. Given a \textit{training set} of input-output pairs, the values of the weight and bias parameters can be ``trained'' so that the model transforms inputs into desired outputs.

In a fully connected neural network, every value in a layer's input is related to every value in its output with a different weight parameter. If layers have large input and output sizes, the network will have many free parameters and training will be computationally expensive. Accordingly, fully connected networks are poorly suited for high-dimensional input data like images: learning complex features requires many large intermediate layers, but this makes training the network computationally infeasible.
\textit{Convolutional neural networks} (CNNs) pose a solution by exploiting spatial structure in their inputs. Whereas a fully connected network treats each pixel in an image independently, a CNN instead learns local features that are detected across the entire input. 
This significantly reduces the number of free parameters and the number of computational operations required to compute the output.

A CNN takes as input a $d$-dimensional grid of input values $\vx_0$. The grid values may be vectors, so $\vx_0$ has shape $(N_1, ..., N_d, n_0)$, where $N_1,...,N_d$ are the lengths of the spatial dimensions and $n_0$ is the length of the input vectors.
The transformation performed by layer $i$ is
\begin{equation}
    \vx_i^{(j)} = \phi \left(\sum_{k} \vw_i^{(k,j)} * \vx_{i-1}^{(k)} + \vb_i^{(j)}\right),
\end{equation}
where $\vx_i$ is a $d$-dimensional grid of output vectors of length $n_i$, $j \in [1, n_i]$ and $k \in [1, n_{i-1}]$ are vector indices, $\vw_i^{(k,j)}$ and $\vb_i^{(j)}$ are $d$-dimensional scalar grids of trainable parameters, $\phi$ is the activation function, and $*$ denotes discrete cross-correlation (colloquially called ``convolution''). CNNs often also include \textit{pooling} layers to downsample the intermediate grids, but we do not use pooling in this paper because our target grids have the same shape as our input grids.

The length $n_i$ of the output vectors is called the number of \textit{filters} of layer $i$. The set of pixels in the input grid contributing to the value of a particular pixel in the output grid comprises the \textit{receptive field} of that pixel. The side length of $\vw_i^{(k,j)}$ is called the \textit{kernel size}. Typically, the kernel size is much smaller than the side length of the input grid (e.g. 3 or 5 pixels in each dimension), so the receptive field of each output pixel is only a small subregion of the input grid. Accordingly, CNNs are local models and require far fewer parameters and computational operations compared to a fully connected neural network of the same output size.

\section{Methods}\label{sec:methods}

\subsection{Combining standard reconstruction and CNNs}\label{sec:cnns_for_recon}

\def\reconwidth{0.24\textwidth}
\begin{figure*}
    \centering
    \begin{subfigure}[b]{\reconwidth}
        \centering
        \includegraphics[width=\textwidth]{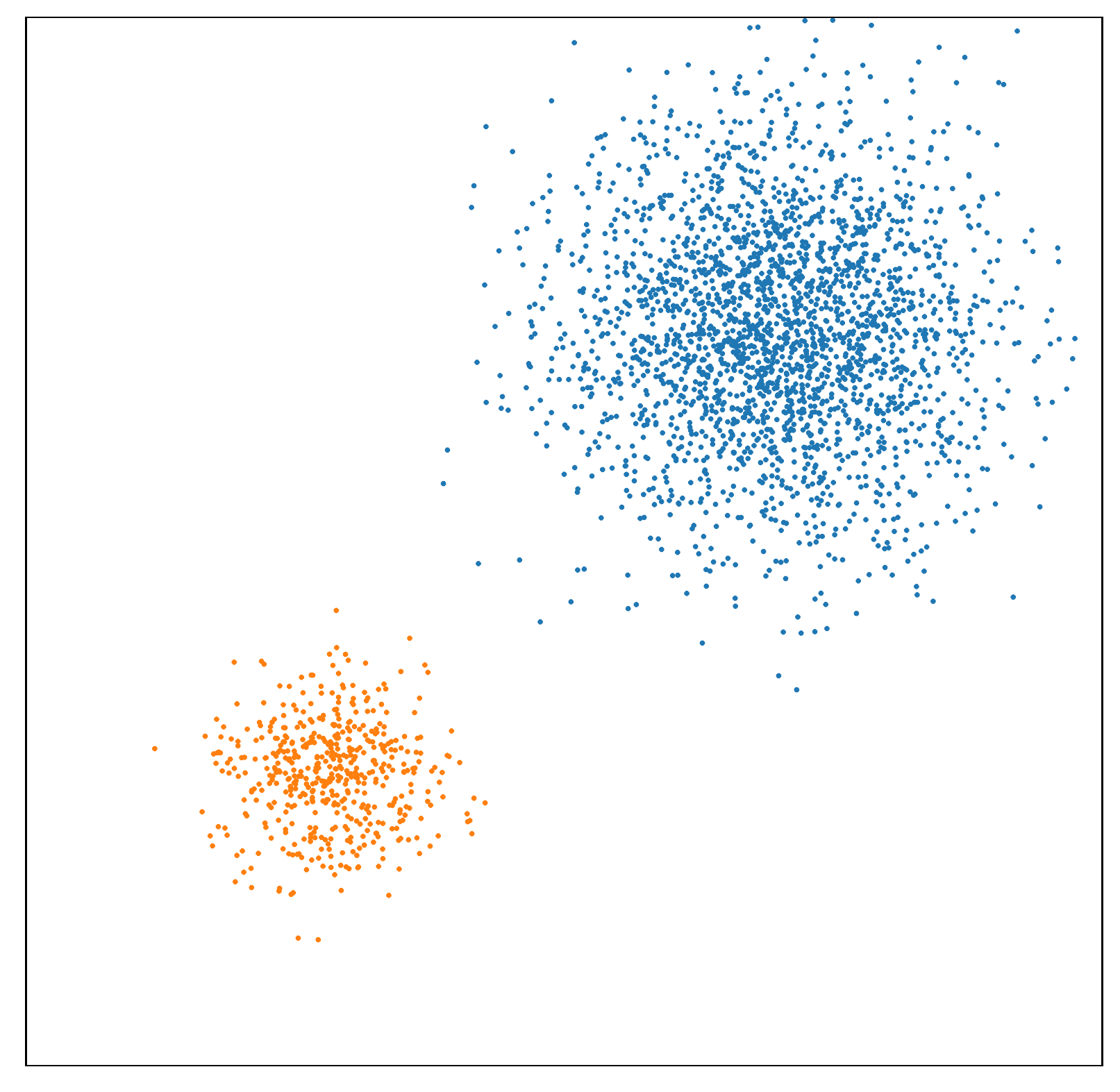}
        \caption{Initial state}
    \end{subfigure}
    \begin{subfigure}[b]{\reconwidth}
        \centering
        \includegraphics[width=\textwidth]{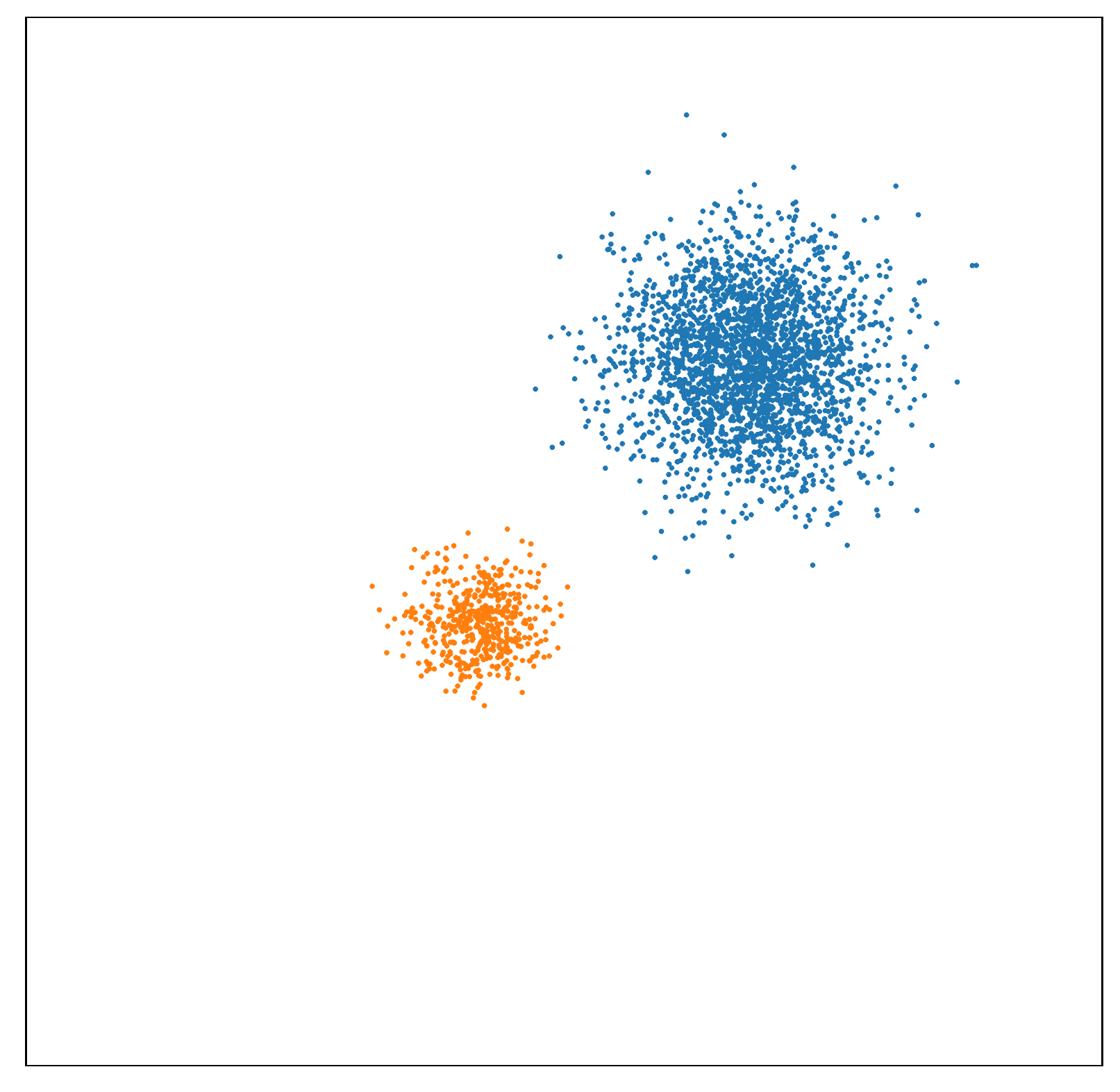}
        \caption{Final state}
    \end{subfigure}
    \begin{subfigure}[b]{\reconwidth}
        \centering
        \includegraphics[width=\textwidth]{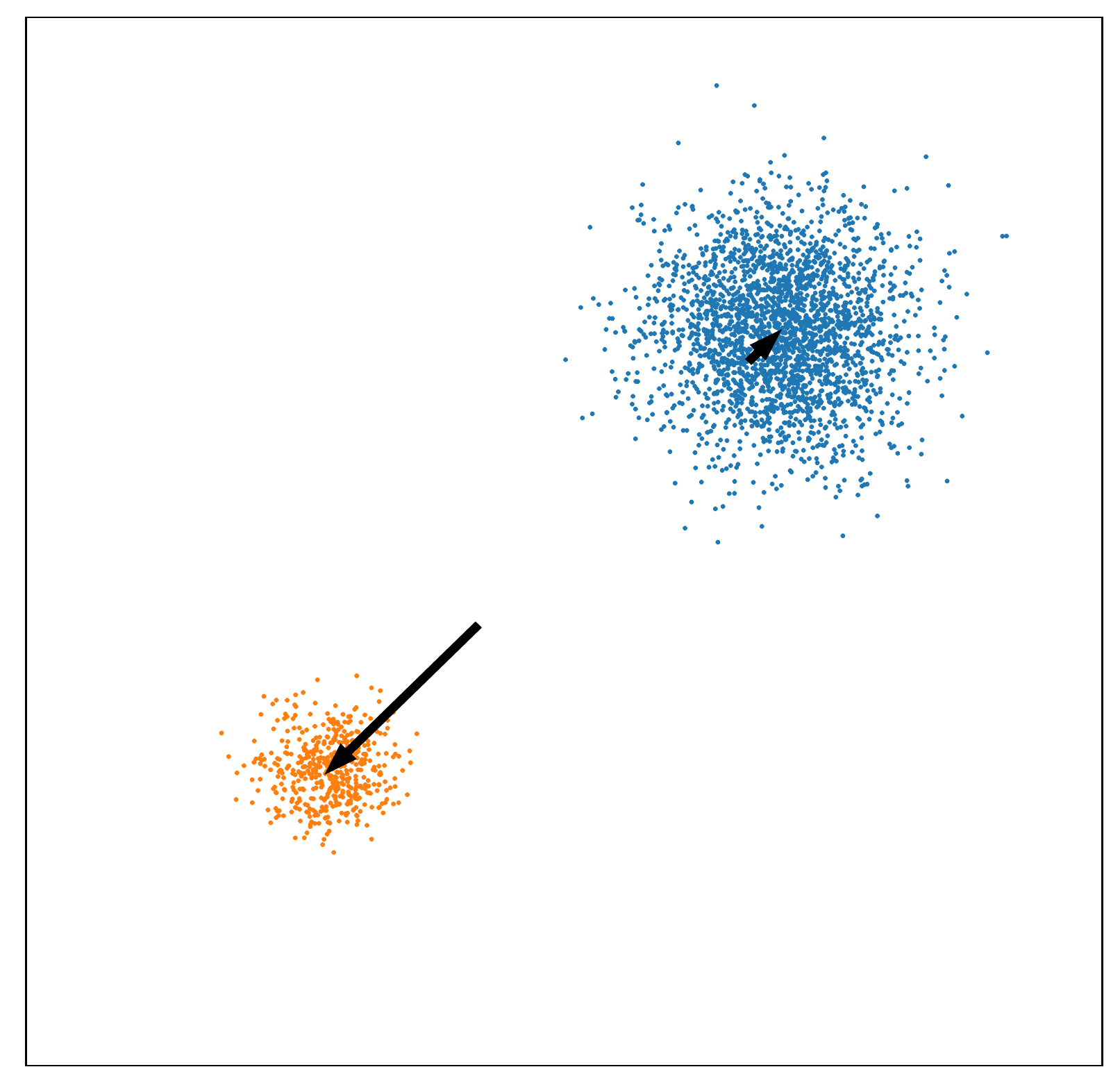}
        \caption{After standard reconstruction}
    \end{subfigure}
    \begin{subfigure}[b]{\reconwidth}
        \centering
        \includegraphics[width=\textwidth]{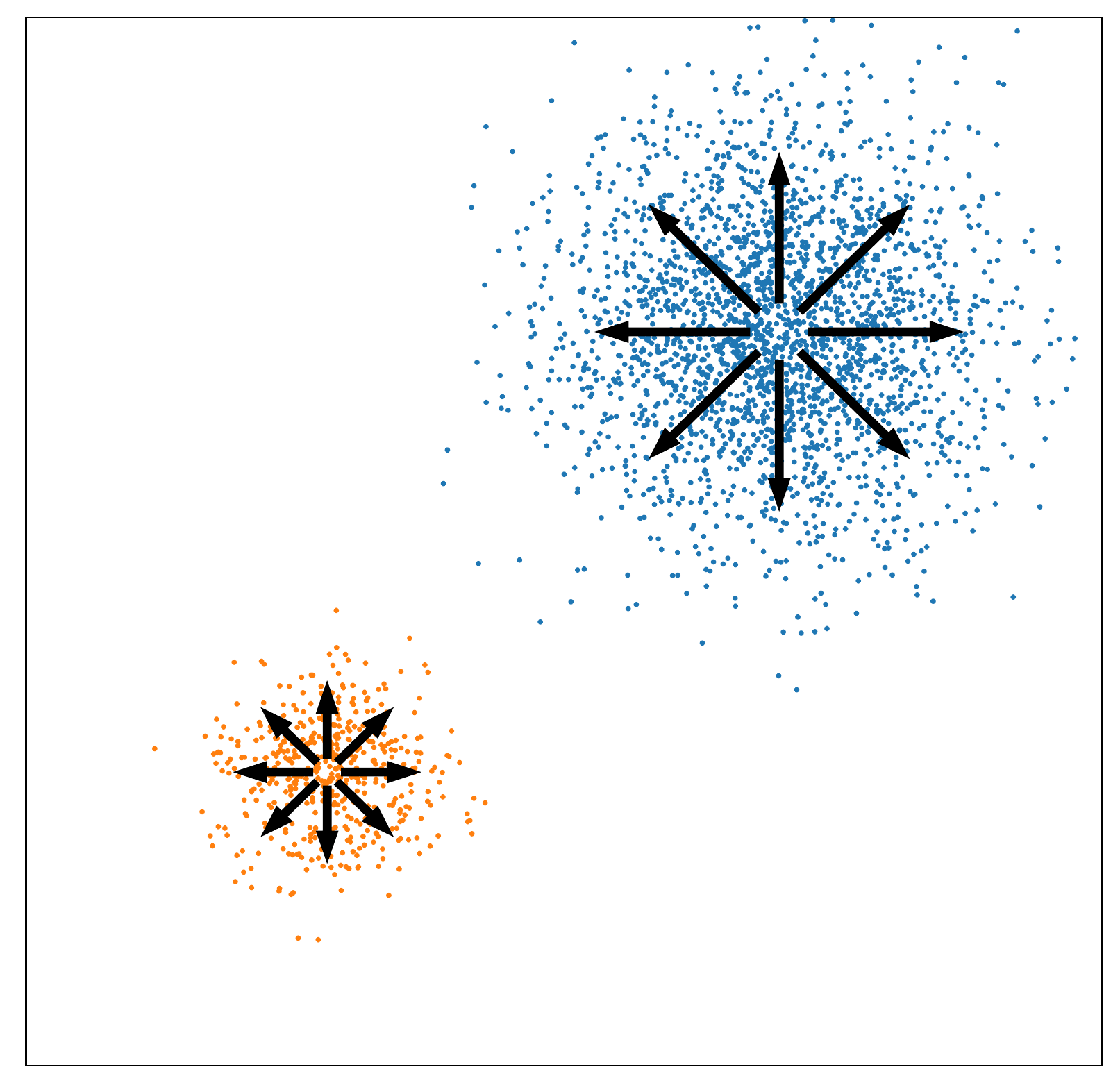}
        \caption{After CNN}
    \end{subfigure}

    \caption{A simplified visualization of our reconstruction method. (a) Two initially overdense regions identified by tracer particles. (b) As the Universe evolves, gravity causes the two regions to collapse in on themselves \textit{and} move closer together. (c) Standard reconstruction inverts gravitational displacements sourced over large scales, corresponding to bulk particle flows but not smaller-scale collapse. (d) After preprocessing with standard reconstruction, the CNN performs a local transformation to invert gravitational collapse at intermediate and small scales.} \label{fig:recon-cartoon}
\end{figure*}

Gravity is a long range force: particle trajectories are determined by gravitational sources distributed over very large scales. In linear theory, under the Zel'dovich approximation, the variance of the displacement of a particle at time $t$ from its initial position $\vq$ satisfies 
\begin{equation}
    \mean{\vec\Psi(\vq, t)^2} = \int \frac{dk}{2 \pi^2} P(k, t),
\end{equation}
where $\mean{\cdot}$ is the ensemble mean, $\vec\Psi$ is the Lagrangian displacement field, and $P(k, t)$ is the power spectrum. In a standard cold dark matter cosmology, this integral attains 50 per cent of its value from wavenumbers smaller than \SI{0.06}{\h\per\mpc}, representing fluctuations over distances greater than \SI{100}{\per\h\mpc}. 

Meanwhile, CNNs are local models: the receptive field of each output pixel covers only a subregion of the entire input grid. Pixels outside the receptive field do not contribute at all, while pixels inside the receptive field but near the edges have far fewer connections to intermediate layers than pixels near the center of the receptive field.

When using a CNN to reconstruct the initial density field, we must ensure that the receptive field of each output pixel contains a sufficient view of both the density that originated at that location \textit{and} the density that sourced the gravitational forces responsible for moving it.
The conflict between the long-range nature of gravitational sources and the inherently local nature of CNNs poses a challenge. For example, with input grid cells of side length $\SI{3.5}{\per\h\mpc}$, an 8-layer CNN with a kernel size of 3 has a receptive field extending just $\SI{28}{\per\h\mpc}$ from the central pixel in each Cartesian direction.
There are various ways to increase the receptive field size, but all approaches come with tradeoffs. We could increase the kernel size or add additional layers, but this increases the number of free parameters and computational complexity of the network -- we would eventually hit practical limits on resources and training time. Alternatively, we could use larger input grid cells or replace the basic convolution operation with a strided convolution \citep[e.g.][]{goodfellow_deep_2016} or dilated convolution \citep{yu_multi-scale_2016}, but this would reduce the resolution of the affected layers, potentially harming the model's predictions on small scales.

In this paper, we propose a solution that avoids the need for large receptive fields: before feeding the late-time density field into the CNN, we apply a preprocessing step to invert bulk gravitational displacements sourced over large scales.
Since gravitational collapse is linear on large scales, reversing the bulk flows is relatively straightforward and is well-handled by existing reconstruction algorithms.
We use \textit{standard reconstruction} \citep{eisenstein_improving_2007,padmanabhan_2_2012} for preprocessing because it is simple and well-studied.
Our two-step reconstruction method is visualized in Figure~\ref{fig:recon-cartoon}.
Using a more sophisticated algorithm for preprocessing would be possible, but the primary goal is to transform the reconstruction problem from long-range to local, not to perform the best possible reconstruction at small and intermediate scales.
As our results in Section~\ref{sec:results} will show, this preprocessing step unlocks the power of CNNs for local transformations, resulting in far better reconstruction compared to either standard reconstruction or a CNN alone.

\subsection{Data set}\label{sec:methods-data}

We developed and evaluated our method using data from \abacussummit, a suite of high-accuracy, publicly available\footnote{\url{https://abacusnbody.org/}} cosmological $N$-body simulations \citep{maksimova_abacussummit_2021,garrison_abacussummit_2021,garrison_abacus_2021,garrison_high-fidelity_2019,garrison_abacus_2018}.
We used the 25 \texttt{base} simulations of the primary cosmology.
Each simulation contains ${\sim}\num{3.3e11}$ dark matter particles in a periodic cube of side length \SI{2}{\giga\parsec\per\h}. The cosmological parameters are based on the Planck 2018 Lambda cold dark matter model \citep{aghanim_planck_2020}. We partitioned the 25 simulations into a training set (15 simulations), validation set (5 simulations), and test set (5 simulations). We used the validation set during the development process to tune the architecture and training parameters, and to monitor our progress. We reserved the test set to evaluate our final performance. Specifically, our splits were:
\begin{itemize}
    \item Test: \texttt{AbacusSummit\_base\_c000\_ph\{000-004\}}.
    \item Training: \texttt{AbacusSummit\_base\_c000\_ph\{005-019\}}
    \item Validation: \texttt{AbacusSummit\_base\_c000\_ph\{020-024\}}
\end{itemize}

For each simulation, we generated the late-time density field $\rho(\vx)$ (i.e. the input to reconstruction) by taking a 3 per cent subsample of particles ($\sim\num{9.9e9}$ particles) at $z=0.5$ and adding them to a $576^3$ comoving grid using periodic triangular-shaped cloud particle distribution scheme. 
We generated two kinds of late-time density fields:
\begin{itemize}
    \item \textit{Real space} density fields, for which we added each particle to the grid at its comoving position at $z=0.5$.
    \item \textit{Redshift space} density fields, for which we applied a redshift distortion to the $z$-component of each particle's comoving position according to its peculiar velocity in that direction. This procedure simulates the positions that would be inferred by an observer located very far away in the $z$ direction.
\end{itemize}
We converted the density field into the \textit{overdensity} field
\begin{equation}
    \delta(\vx) \equiv \frac{\rho(\vx) - \bar{\rho}}{\bar{\rho}},
\end{equation}
where $\bar{\rho}$ is the mean density.

We paired each late-time overdensity grid with a corresponding initial overdensity grid (i.e. the target output of reconstruction) at $z=99$. We used the \abacussummit\ initial condition grids, which -- unlike our late-time density fields -- were generated directly from the power spectrum rather than by adding discrete particles to a grid. We could alternatively have generated the initial condition grids from a discrete subsample of particles: we do not think this would make a significant difference to our results. Note that the initial overdensity grid is always in real space, even when the late-time overdensity grid has redshift distortions.

We note that each of the 25 density field pairs is a substantial data set in its own right, containing $576^3 \approx \num{2e8}$ points in each late-time and initial overdensity grid.
We used 15 simulations in our training set because they were readily available, but this amount of training data is overkill: we saw very similar results when we trained our CNN on just one simulation instead of 15.
If future applications of our method require generating fresh training data, then those applications can likely get away with a far smaller training set than we used.

\subsection{Preprocessing}\label{sec:methods-recon}

As discussed in Section~\ref{sec:cnns_for_recon}, we applied standard reconstruction \citep{eisenstein_improving_2007} before feeding the late-time density field into our CNN.
In real space, we first smoothed the density field with an isotropic 3D Gaussian filter of comoving width $\sigma = \SI{10}{\per\h\mpc}$ to filter out small-scale nonlinear perturbations. Then we computed the Lagrangian displacement field $\vec\Psi(\vx)$ by solving the linear perturbation theory continuity equation
\begin{equation}
    \nabla \cdot \vec\Psi(\vx) = - \delta(\vx)
\end{equation}
assuming that $\vec\Psi(\vx)$ is irrotational.
We generated two new overdensity fields $\delta_d(\vx)$ and $\delta_r(\vx)$ by displacing the particles and a set of $10^{10}$ random particles by $-\vec\Psi(\vx)$, respectively. We drew the random particle positions from a uniform distribution and set their total mass equal to the total mass of the real particles. Finally, we computed the (partially) reconstructed overdensity field by subtracting the two displaced overdensity fields:
\begin{equation}
    \delta_\text{rec} \equiv \delta_d(\vx) - \delta_r(\vx).
\end{equation}
In redshift space, we followed the same procedure except for the following modifications \citep{padmanabhan_2_2012,seo_modeling_2016}. When computing $\vec\Psi(\vx)$, we solved
\begin{equation}
    \nabla \cdot \vec\Psi(\vx) + f \frac{\partial}{\partial z} \left( \vec\Psi(\vx) \cdot \hat{\vz} \right) = - \delta(\vx),
\end{equation}
where $\hat{\vz}$ is the direction of redshift distortions and $f \equiv \text{d} \, \ln{D} / \text{d} \ln a$ is the linear growth rate, where $D(a)$ is the linear growth function as a function of scale factor. We used $f = 0.759$ at $z=0.5$ from the simulation cosmology.
When computing $\delta_d(\vx)$, we displaced the particles by an additional $- f (\vec\Psi(\vx) \cdot \hat{\vz}) \hat{\vz}$ to partially correct redshift distortions. We did not apply this additional displacement to the random particles.

As we will discuss further in Section~\ref{sec:results-redshift}, we also generated transformations of the partially-reconstructed overdensity field and considered CNN architectures that used these transformations as additional inputs. These included smoothed copies of the partially-reconstructed overdensity field at different smoothing scales. We computed these smoothed grids by convolving with isotropic 3D Gaussian filters of comoving width $L, 2L, 4L, ...$, where $L \approx \SI{3.5}{\per\h\mpc}$ is the width of one cell. We also generated first and second order gradients of these smoothed grids. We will discuss the motivations and results of using these additional inputs in Section~\ref{sec:results-redshift}.

We pre-computed the late-time overdensity field and performed standard reconstruction for each of the 25 simulations in our training, validation, and test sets so that we did not have to perform these time-intensive computations many times when training our CNNs. However, we generated any additional transformations of these grids (smoothed copies and gradients) on-the-fly during training since pre-computing all of the transformed grids would have consumed significant disk space.

\subsection{CNN architecture}\label{sec:methods-cnn}

Our CNN architecture is shown in Table~\ref{tab:cnn}.
The input is an $n^3 \times d$ grid, where the first 3 dimensions are spatial and the fourth is the number of scalar inputs per grid cell. In the basic case we input just the overdensity field, so that $d=1$, but 
in Section~\ref{sec:results-redshift} we will have $d>1$ when we include additional transformations of the overdensity field.
The input grid is passed through 8 intermediate convolutional layers, each with 32 filters and hyperbolic tangent activation function, followed by a final convolutional layer with a single feature and no activation function. All convolutional layers have kernel size 3.
The output is an $(n-18)^3$ scalar grid representing the reconstructed initial overdensity field; cells within 9 pixels of the boundary are not reconstructed because their receptive fields are partly outside the input grid. If we assume  periodic boundary conditions, we can reconstruct the entire grid by periodically extending the input by 9 pixels on each side.

We chose our final CNN architecture by tuning the architectural specifications during development.
We varied the number of intermediate convolutional layers between 1 and 20 and the number of features of each intermediate layer between 8 and 64. In general, reconstruction performance improved as the number of layers and number of features increased. We settled on 8 intermediate layers with 32 features per layer because the improvements had mostly saturated by that point. Our goal was to maximize the final performance rather than minimize the training time or number of free parameters in the CNN. Indeed, once we had preprocessed the late-time overdensity field with standard reconstruction, even a CNN with just one intermediate layer of 8 features significantly improved the reconstruction (although not as much as our more sophisticated final CNN architecture).

\begin{table}
    \centering
    \caption{Our CNN architecture. Each convolutional layer reduces the size of the output grid by one pixel on each spatial boundary. Assuming periodic boundary conditions, we can reconstruct an entire cosmological grid by periodically extending it by 9 pixels on each side.}
    \label{tab:cnn}
    \renewcommand{\arraystretch}{1.3}
    \begin{tabular}{ccccc}
     \hline
     Layer & Kernel size & Num. filters & Activation & Output Shape\\
     \hline
     \hline
     input & - & -  & - & $n^3 \times d$ \\
     \hline
     1 & 3 & 32 & tanh & $(n-2)^3 \times 32$ \\ 
     \hline
     2 & 3 & 32 & tanh & $(n-4)^3 \times 32$ \\
     \hline
     \vdots & \vdots & \vdots & \vdots & \vdots \\
     \hline
     8 & 3 & 32 & tanh & $(n-16)^3 \times 32$ \\
     \hline
     9 & 3 & 1 & - & $(n-18)^3 \times 1$ \\
     \hline
    \end{tabular}
\end{table}

We also experimented with residual connections \citep{he_deep_2015} and dilated convolutions \citep{yu_multi-scale_2016}. Residual connections make it easier to optimize very deep CNNs by reformulating layers as residual functions to the identity map.  We did not see any improvement from residual connections even with our deepest models, but residual connections are commonly used in much deeper models (e.g. 50+ layers). Dilated convolutions increase the receptive field at the expense of small-scale resolution. We found that CNNs with dilated convolutions performed worse than CNNs of the same depth with non-dilated convolutions.

\subsection{Training}\label{sec:methods-training}

We must \textit{train} our CNN to produce the desired mapping from input grid to output grid. This involves finding the values of the convolutional kernel and bias parameters that best reproduce the mapping between input grid and target grid in the training set.

\subsubsection{Optimization objective}

The canonical objective for regression is to minimize the mean squared error loss function
\begin{equation}\label{eq:loss-mse}
    L_\text{MSE} = \frac{1}{N_\text{sims}} \sum_i \frac{1}{N_\text{cells}} \sum_{\vx} \left( \delta^{(i)}_\text{recon}(\vx) - \delta^{(i)}_\text{init}(\vx) \right)^2,
\end{equation}
where $\delta^{(i)}_\text{recon}(\vx)$ and $\delta^{(i)}_\text{init}(\vx)$ are the predicted and true initial overdensity grids of simulation $i$; $N_\text{cells}$ is the number of grid cells per simulation; and $N_\text{sims}$ is the number of simulations in the training set. The values of the CNN parameters that minimize the mean squared error are the maximum likelihood solution under the assumption that the cell-by-cell predictions of the ``true'' model differ from the true initial conditions by a zero-mean Gaussian with constant variance.

We can interpret the mean squared error in Fourier space using Parseval's theorem\footnote{Here we are assuming that $\delta^{(i)}_\text{recon}(\vx)$ and $\delta^{(i)}_\text{init}(\vx)$ are bandlimited with maximum wavenumber less than the Nyquist wavenumber of the grid.} as
\begin{equation}\label{eq:loss-fourier}
    L_\text{MSE} =  \frac{1}{N_\text{sims}} \sum_i \frac{1}{V^2} \sum_{\vk} \left| \ft{\delta}^{(i)}_\text{recon}(\vk) - \ft{\delta}^{(i)}_\text{init}(\vk) \right|^2,
\end{equation}
where $V$ is the volume of each simulation and where $\tilde{f}(\vk) \defeq \int_V d\vx \, e^{-i \vk \cdot \vx} f(\vx)$ denotes the Fourier series coefficient at wavevector $\vk$ of a periodic function $f(\vx)$ on $V$. Thus, minimizing the mean squared error is equivalent to minimizing the squared difference between Fourier coefficients on a mode-by-mode basis.

We used the mean squared error loss function to train our CNNs.
In machine learning it is common to add additional regularization terms to the loss function to prevent overfitting or encourage desired properties of the solution. We will return to this in Section~\ref{sec:results-redshift}.

\subsubsection{Optimization algorithm}

We trained our CNNs using stochastic gradient descent (SGD) with momentum \citep{polyak_methods_1964}. Stochastic gradient methods, including SGD and more recently proposed variants like Adam \citep{kingma_adam_2015}, are the most popular algorithms for training deep neural networks. These algorithms iteratively reduce the training loss by computing its gradient with respect to the model parameters and taking a small step in the direction of maximum decrease. Rather than compute the full loss at each training step, which would require iterating through the entire training set, each step is made using a stochastic estimate of the gradient from a random subset of the training set.

During training, we randomly selected a simulation from the training set and loaded the input grid and its corresponding target grid. In cases where we included transformations of the input grid (see Section~\ref{sec:results-redshift}), we performed these transformations on the fly. This yielded a $576^3 \times d$ input grid and corresponding $576^3$ target grid. Due to GPU memory constraints, we then randomly took a $146^3 \times d$ slice of the input grid and corresponding $128^3$ slice of the target grid, computed a stochastic estimate of the gradient of the loss, and updated the model parameters accordingly.
Since loading the data for each simulation is much slower than a training step -- especially when computing on-the-fly transformations of that data -- we performed 20 training steps each time we loaded a simulation, each with a different randomly selected slice of that simulation. Known as ``data echoing,'' this technique has been demonstrated to substantially speed up neural network training when limited by input loading and preprocessing time, even though it technically violates the independence of gradient estimates assumed by stochastic gradient methods \citep{choi_faster_2020}.

During training, we re-scaled the input and target grids such that their spatial variances were near unity.
We used a learning rate of $0.01$ and momentum parameter $0.99$, and trained all models for \num{10000} steps. We tuned these values manually to maximize final performance. We note that it is possible to get results that are almost as good with far fewer training steps.
We tried using Adam instead of SGD, but we didn't see a significant improvement.
We monitored for overfitting by periodically computing the loss on the training set and the loss on a validation simulation over the course of training. The two matched very closely for all of our training runs, indicating that overfitting was not an issue.

\subsubsection{Polyak averaging}

Since our training algorithm is stochastic (i.e. each update only considers a random subgrid of a random simulation), the CNN's parameters tend to fluctuate even at the end of training. In particular, we observed that the overall scale of the predicted output grid continued to fluctuate even as the correlation between the predicted and target grids converged. We believe this is due to chance selections of particularly over- or under-dense subgrids, which shifted the overall normalization of the entire model. To stabilize the final model, we maintained an exponential moving average of the model parameters over training,
\begin{equation}
    \bar{\vec\theta}_{n} = \alpha \vec\theta_n + (1-\alpha) \bar{\vec\theta}_{n-1},
\end{equation}
where $\vec\theta_n$ denotes the vector of all CNN parameters parameters at step $n$, $\bar{\vec\theta}_{n}$ is its exponential moving average, and $\bar{\vec\theta}_0 \defeq \vec\theta_0$. We used $\alpha = 0.01$. We take as our final model $\bar{\vec\theta}_{N}$, the exponential moving average at the final step. This is a variant of Polyak averaging \citep{polyak_acceleration_1992} commonly used for training neural networks.

\subsection{Implementation and code availability}

We implemented our model and training code in \texttt{jax} \citep{bradbury_jax_2018} using the \texttt{flax} \citep{heek_flax_2020} and \texttt{optax} \citep{hessel_optax_2020} libraries. We initialized bias parameters to zero and kernel parameters using the \texttt{jax.nn.initializers.lecun\_normal()} initialization scheme\footnote{\url{https://jax.readthedocs.io/en/latest/\_autosummary/jax.nn.initializers.lecun\_normal.html}}. Our code is publicly available.\footnote{\url{https://github.com/cshallue/recon-cnn}}

We trained each model on a NVIDIA GeForce GTX 1080 Ti GPU. Training time varied based on the complexity of the CNN inputs. For models using a scalar input grid read directly from disk, training took approximately 4 hours. For models using smoothed versions of the input grid at different scales (see Section~\ref{sec:results-redshift}), training took around 12 hours. The increase in training time was mostly due to smoothing the input grid on-the-fly. Note that our focus was achieving the best performance within a reasonable training time, rather than minimizing the training time. Using a smaller version of our CNN and/or training for fewer steps still significantly improves on standard reconstruction, although not quite to the extent of our final setup.

\section{Results}\label{sec:results}

First we shall describe the metrics and notation that we will use throughout this section.

The 2-point correlation function of a homogeneous overdensity field $\delta(\vx)$ is 
\begin{equation}\label{eq:xi}
    \xi(\vr) \equiv \mean{\delta(\vx) \delta(\vx + \vr)},
\end{equation}
where $\vx$ is any point, $\vr$ is a separation vector, and $\mean{\cdot}$ denotes the ensemble mean. We computed $\xi(\vr)$ by replacing the ensemble mean with a spatial mean over $\vx$, which turns equation~\eqref{eq:xi} into a spatial autocorrelation.

The power spectrum of $\delta(\vx)$ is
\begin{equation}
    P(\vk) \equiv \mean{|\tilde{\delta}(\vk)|^2}.
\end{equation}

In redshift space, in which our fields are not isotropic, we expanded $\xi(\vr)$ and $P(\vk)$ in Legendre polynomials:
\begin{equation}
    \xi(\vr) = \sum_{\ell = 0}^\infty \xi_\ell(r) p_\ell(\cos\theta), \quad  P(\vk) = \sum_{\ell = 0}^\infty P_\ell(k) p_\ell(\cos\theta)
\end{equation}
where $p_\ell$ is the $\ell$-th Legendre polynomial and $\theta$ is the angle between the redshift direction and $\vr$ or $\vk$.

We use the following metrics to compare a reconstructed initial overdensity field $\delta_\text{recon}(\vx)$ to the true initial overdensity field $\delta_\text{init}(\vx)$.
The \textit{correlation coefficient}  between the reconstructed and actual initial fields is
\begin{equation}
    C(\vk) \equiv
    \frac{\mean{\tilde{\delta}_\text{recon}(\vec{k}) \tilde{\delta}_\text{init}(\vec{k})^*}}{\sqrt{P_\text{recon}(\vec{k}) P_\text{init}(\vec{k})}},
\end{equation}
where $^*$ denotes complex conjugation, and the \textit{transfer function} is
\begin{equation}
    T(\vk) \equiv \sqrt{\frac{P_\text{recon}(\vk)}{P_\text{init}(\vk)}}.
\end{equation}
Intuitively, the correlation coefficient measures the linear correlation between the fields as a function of wavevector, whereas the transfer function measures the difference in magnitude.
Note that these metrics are not explicitly optimized by the training algorithm, which minimizes a loss function consisting of the mean squared error plus regularization terms.
The correlation coefficient and transfer function give more insight than the mean squared error because they decompose correlation and magnitude.
Perfect reconstruction (i.e. $\delta_\text{recon}(\vec{x}) = \delta_\text{init}(\vec{x})$) is equivalent to $C(\vk) = T(\vk) = 1$.

We computed $C(\vk)$ and $T(\vk)$ as functions of wavenumber $k$ by replacing ensemble means with spatial averages over bins in $|\vk|$. Thus, the resulting functions $C(k)$ and $T(k)$ are averaged over all directions.

We computed all of these metrics using the fast Fourier transform (FFT). For fields that were generated from discrete particles, we computed their Fourier coefficients by adding the particles to a $1152^3$ grid using triangular-shaped-cloud (TSC) particle distribution scheme, performing the FFT, and deconvolving with the Fourier coefficients of the TSC window function. We used a grid size of $1152^3$, rather than $576^3$, to avoid aliasing effects when computing metrics out to $k=0.9$, which is close to the Nyquist frequency of a $576^3$ grid.
Our initial density grids were generated analytically and explicitly bandlimited, so they do not suffer from aliasing or discreteness effects. Our CNN outputs were trained to reproduce the initial density grids, so we treat them the same way.

All results in this section are computed over the entire $\SI{8}{\giga\parsec\cubed\per\h\cubed}$ volume of a simulation from the test set. This test simulation was not seen by the model during training, nor was it used to refine the model architecture or training parameters. It used the same cosmology as the training simulations, but different initial conditions. We only show results from one of our 5 test set simulations because the others give very similar results.

\def\subfigwidth{0.48\textwidth}
\begin{figure*}
    \centering
    \includegraphics[width=\subfigwidth]{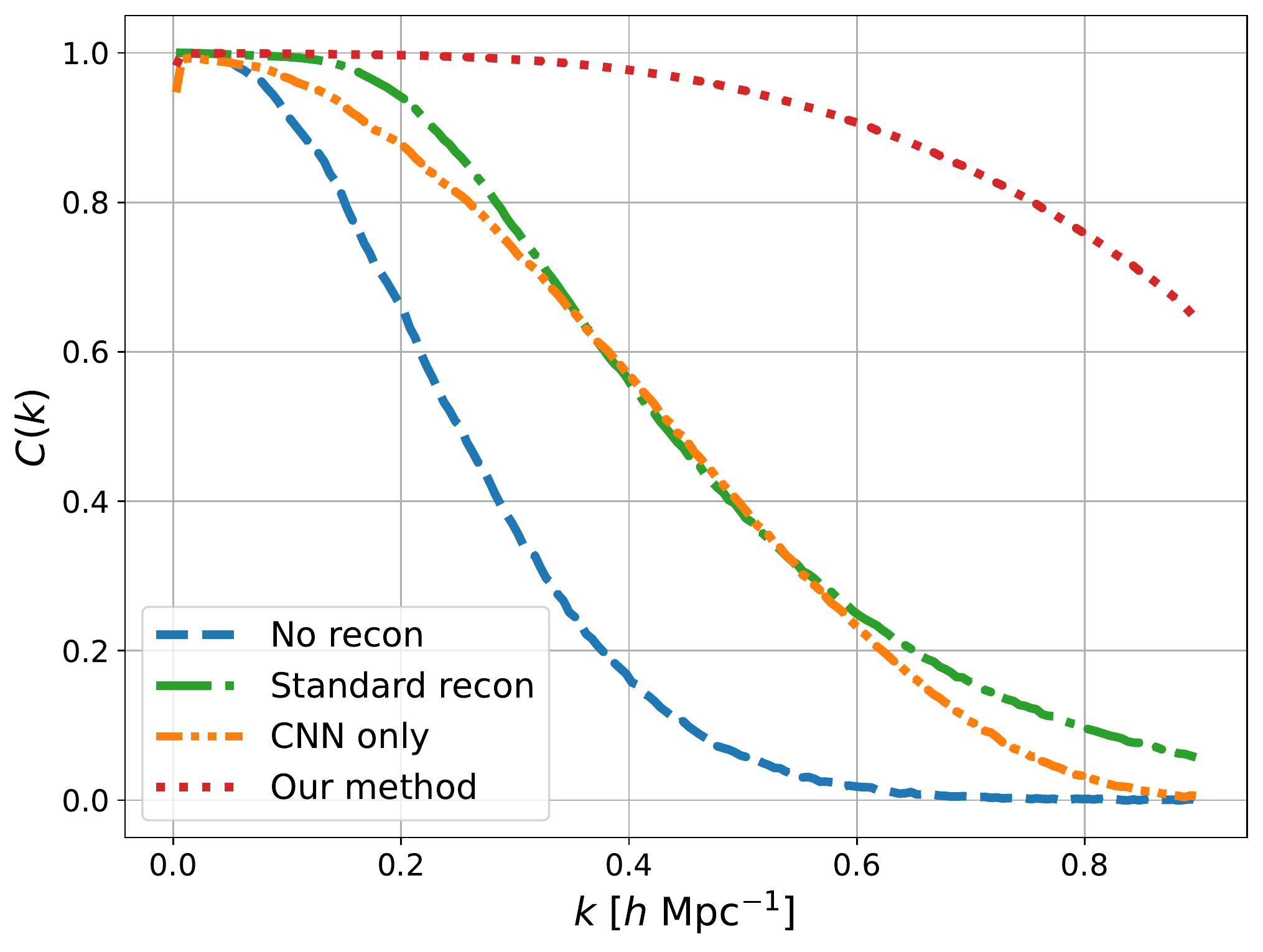}
    \hfill
    \includegraphics[width=\subfigwidth]{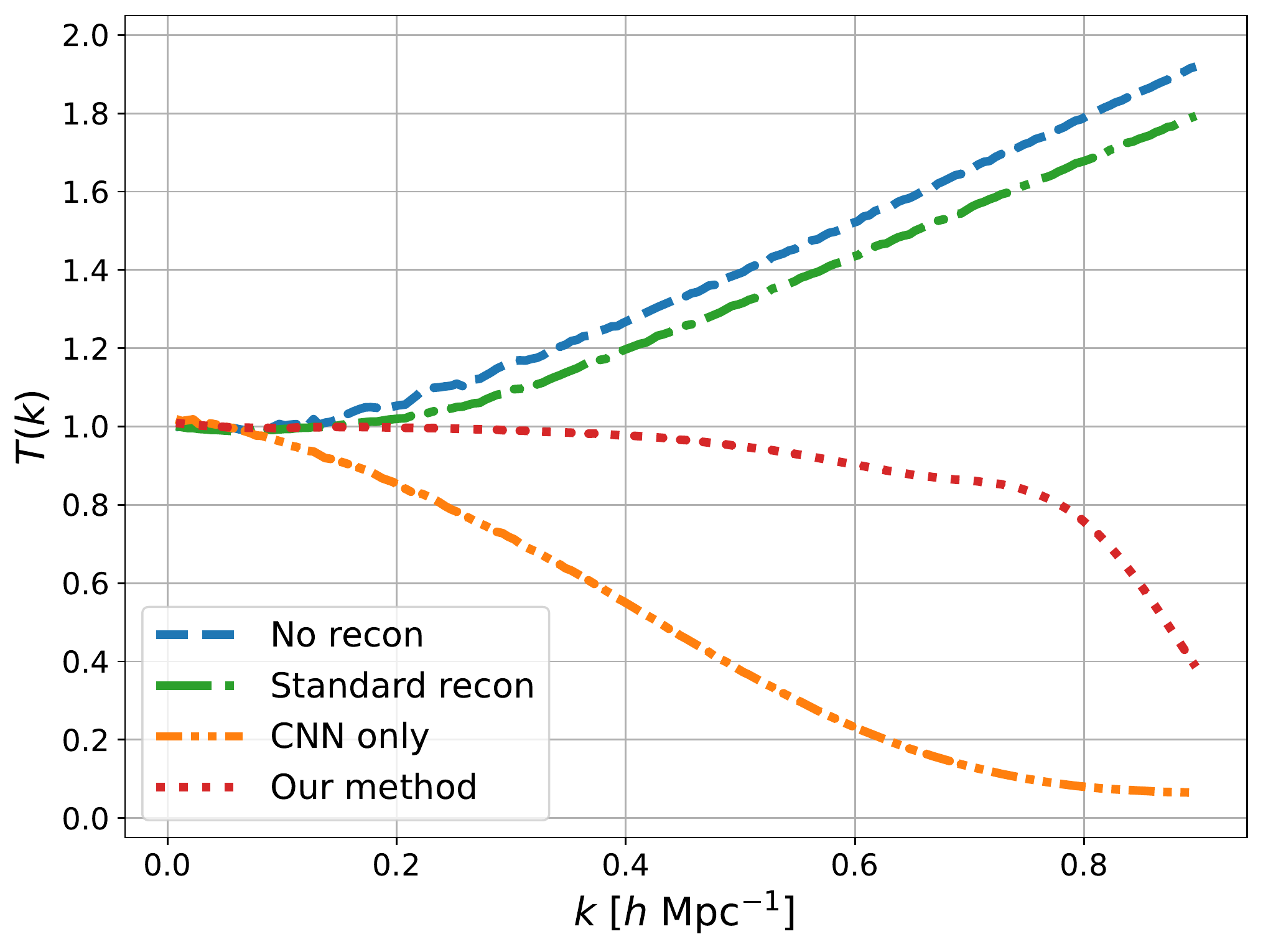}
    \caption{Correlation coefficient and transfer function with respect to the true initial overdensity field ($z=99$) for real-space reconstruction. {\red ``No recon''} is the overdensity field at $z=0.5$. {\red ``Standard recon''} is the initial overdensity field produced by standard reconstruction. {\red ``CNN only''} is the initial overdensity field produced by our CNN when trained on the raw overdensity field. {\red ``Our method''} is the initial overdensity field produced by our CNN when trained on the output of standard reconstruction. To normalize the scales of the transfer function, the fields at $z=99$ have been multiplied by $D(z=0.5) / D(z=99)$.}\label{fig:corrs_real}

    \includegraphics[width=\subfigwidth]{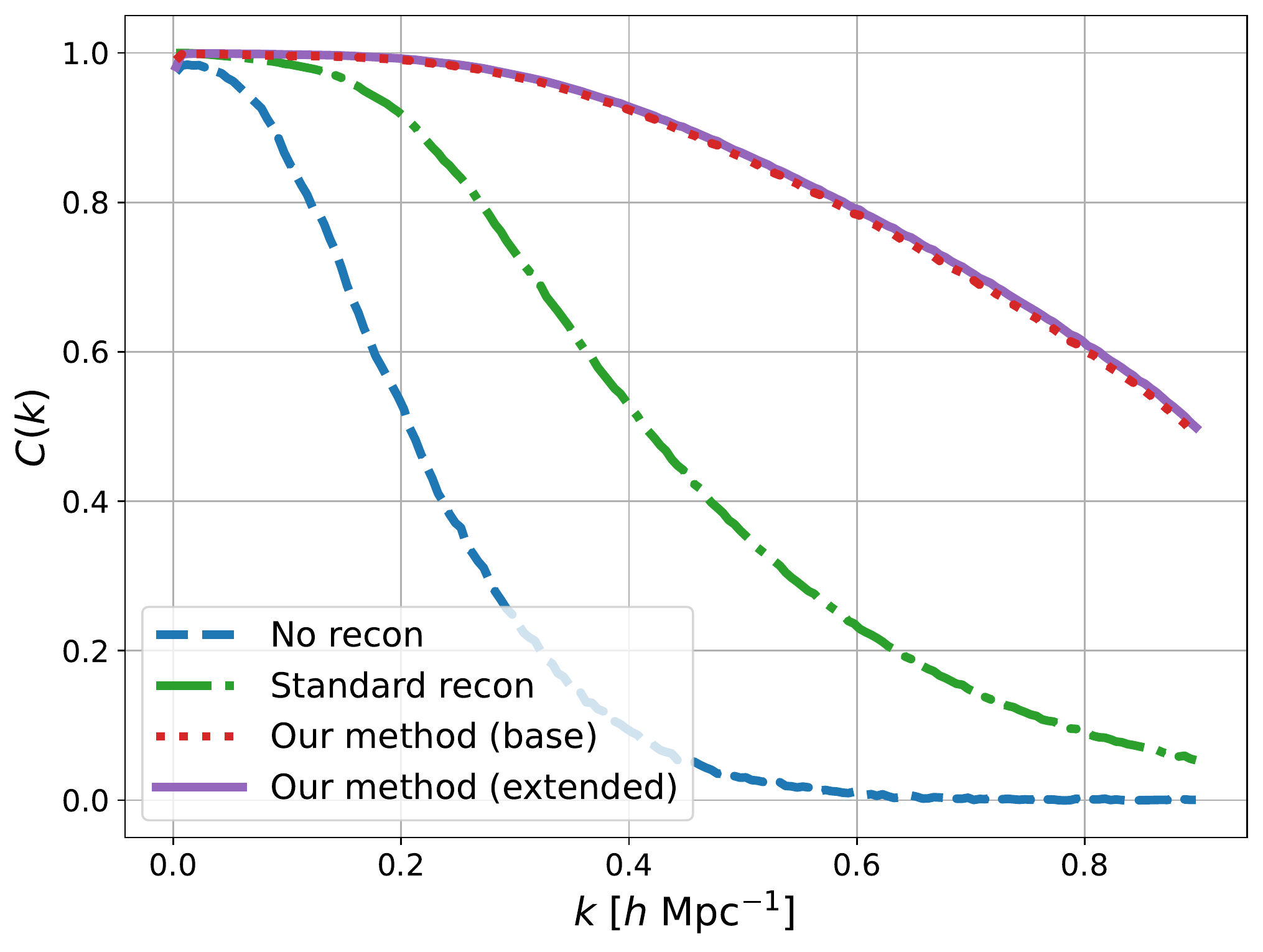}
    \hfill
    \includegraphics[width=\subfigwidth]{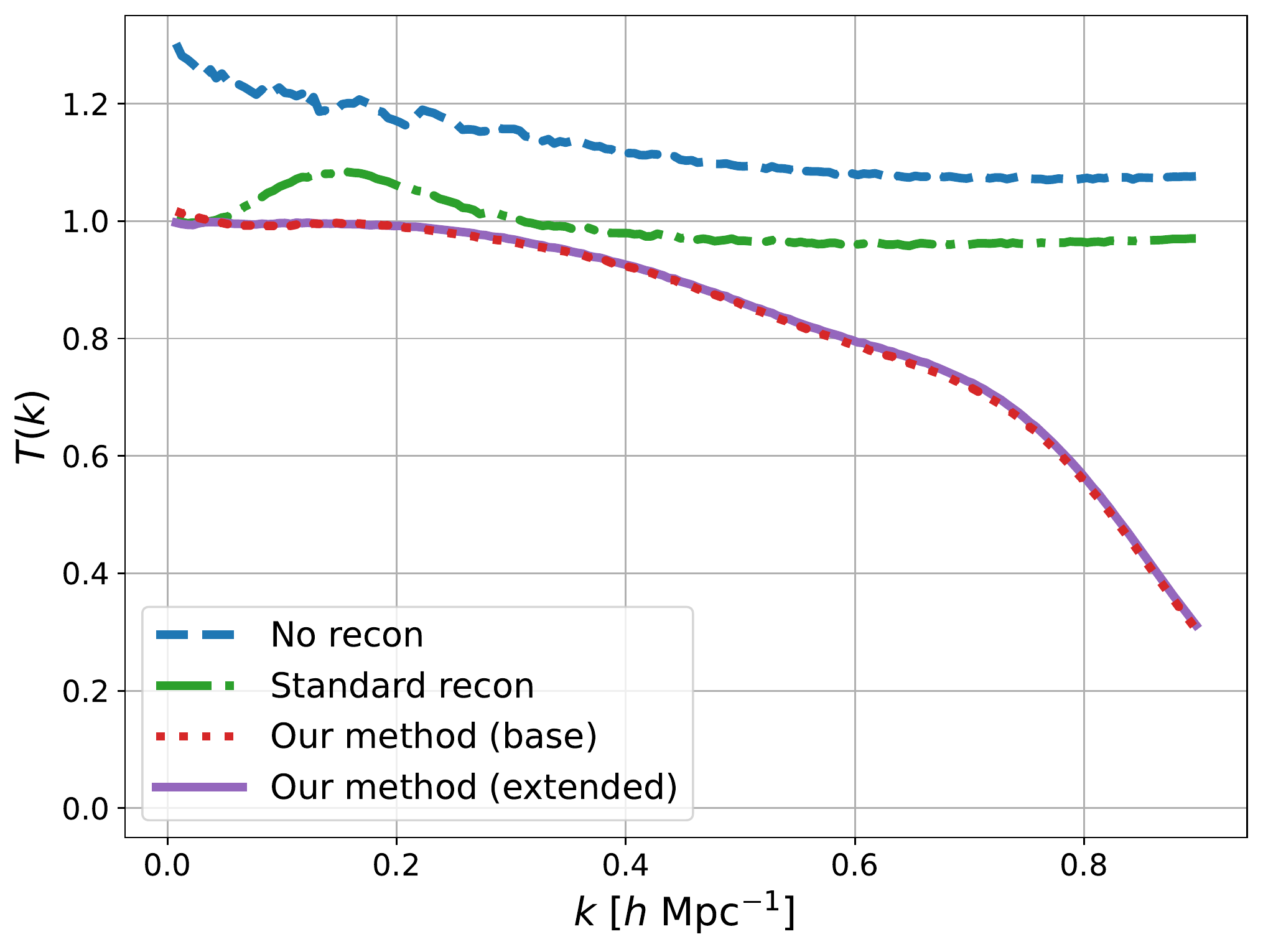}

    \caption{Correlation coefficient and transfer function with respect to the true initial overdensity field ($z=99$) for redshift space reconstruction. {\red ``No recon''} is the redshift-space overdensity field at $z=0.5$. {\red ``Standard recon''} is the initial overdensity field produced by standard reconstruction. {\red ``Our method''} is the initial overdensity field produced by our CNN when trained on the output of standard reconstruction; {\red the extended method} includes the additional inputs and regularization term described in Section~\ref{sec:results-redshift}. To normalize the scales of the transfer function, the fields at $z=99$ have been multiplied by $D(z=0.5) / D(z=99)$.}\label{fig:corrs_redshift}
\end{figure*}

\begin{figure*}
    \centering
    \includegraphics[width=\subfigwidth]{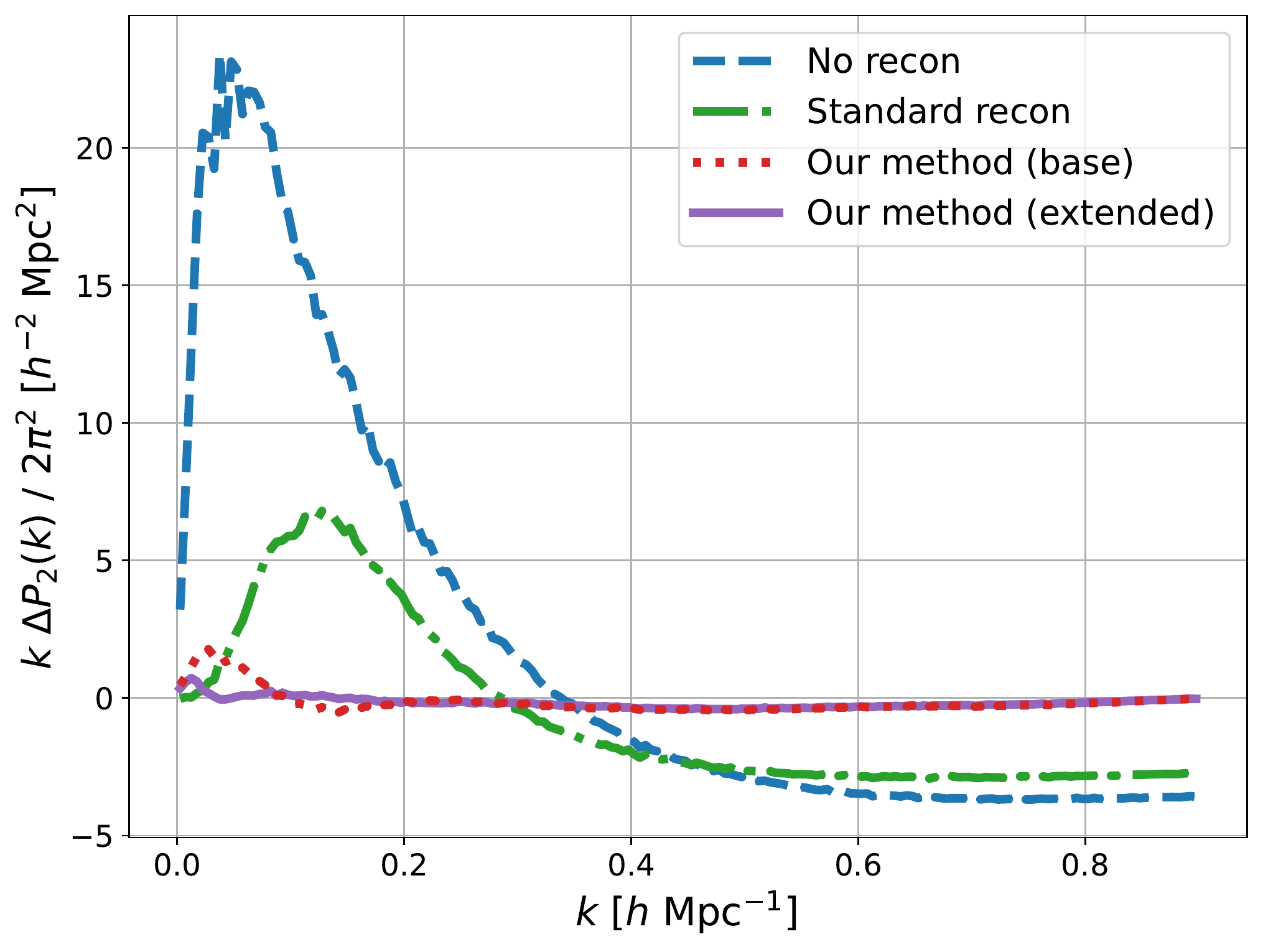}
    \hfill
    \includegraphics[width=\subfigwidth]{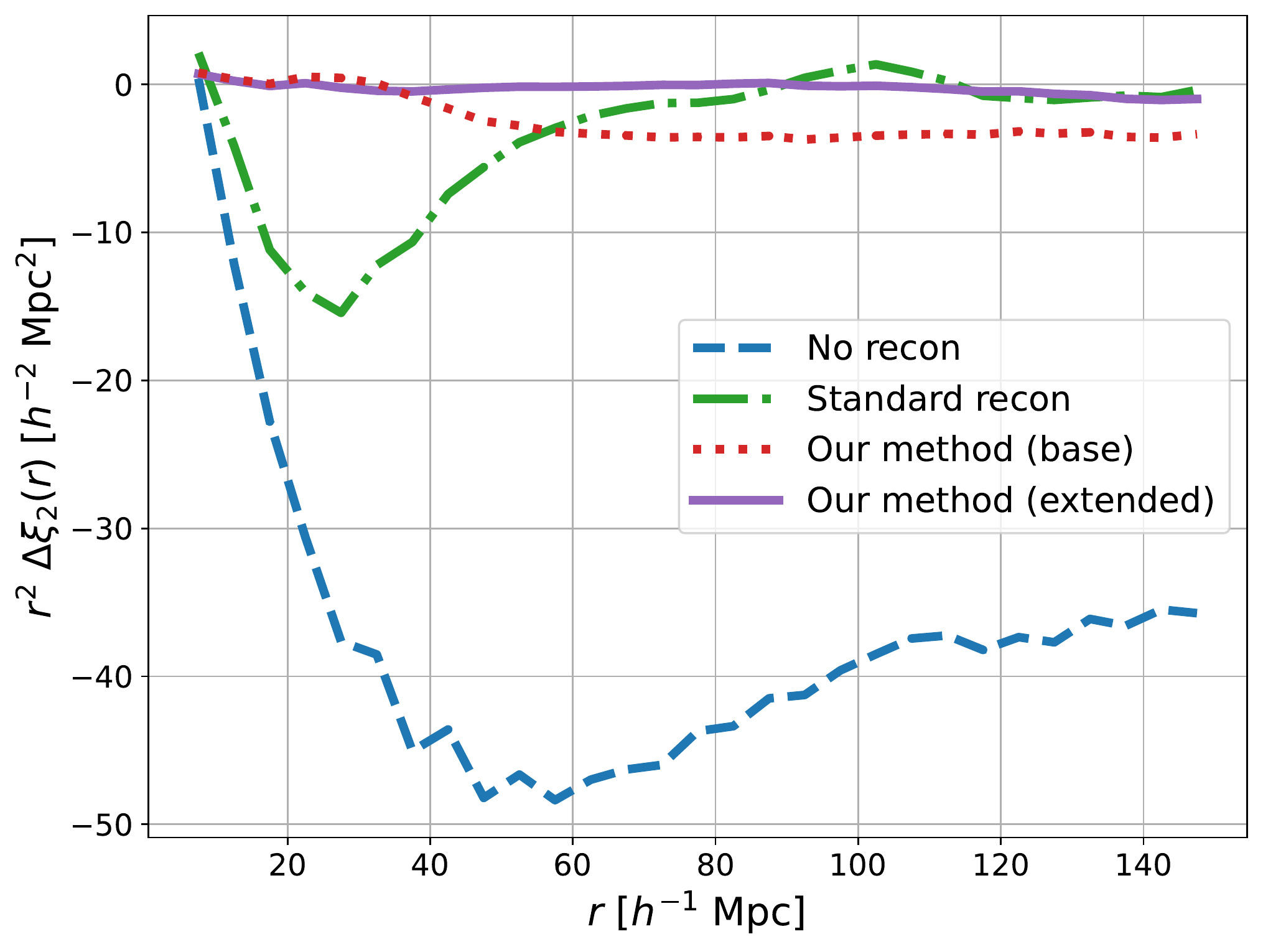}

    \caption{Quadrupole of the power spectrum $P(\vk)$ and 2-point correlation function $\xi(\vr)$ relative to the real-space initial overdensity field at $z=99$. {\red ``No recon''} is the redshift-space overdensity field at $z=0.5$. {\red ``Standard recon''} is the initial overdensity field produced by standard reconstruction. {\red ``Our method''} is the initial overdensity field produced by our CNN when trained on the output of standard reconstruction; {\red the extended method} includes the additional inputs and regularization term described in Section~\ref{sec:results-redshift}. To normalize the scales of the transfer function, the fields at $z=99$ have been multiplied by $D(z=0.5) / D(z=99)$.}\label{fig:quadrupole}
\end{figure*}

\subsection{Real space reconstruction}\label{sec:results-real}

Figure~\ref{fig:corrs_real} compares three methods for reconstructing the initial overdensity field from the late-time real-space overdensity field:
\begin{enumerate}
    \item standard reconstruction, as described in Section~\ref{sec:methods-recon};
    \item our CNN, when inputting the late-time overdensity;
    \item our CNN, when inputting the late-time overdensity after preprocessing with standard reconstruction.
\end{enumerate}

If our CNN is trained directly on the late-time overdensity field \textit{without} applying standard reconstruction first (i.e. method (ii)) then it fails to improve upon standard reconstruction in terms of either the correlation coefficient or transfer function.
As we argued in Section~\ref{sec:cnns_for_recon}, CNNs are not well suited to transforming the late-time density directly into the initial density because the relationship between the two fields is not local.
Our results are similar to those of \citet{mao_baryon_2020}, who also trained a CNN for this task. The biggest difference between our CNN and that of \citet{mao_baryon_2020} is their use of strided convolutions, which are equivalent to regular convolutions with downsampling \citep[][]{goodfellow_deep_2016}. Compared to regular convolutions with the same number of parameters, strided convolutions increase the receptive field while reducing the resolution of the input.
Accordingly, our CNN has a much smaller receptive field, but maintains full resolution in every layer. Our CNN is also more efficient to train because neighboring output cells share intermediate computations (unlike for strided convolutions).

The method we propose in this paper (i.e. method (iii)) significantly outperforms both other methods at all scales in terms of the correlation coefficient and transfer function.
Our method produces high-fidelity reconstruction ($C(k) > 0.95$) for $k \leq \SI{0.5}{\h\per\mpc}$ versus $k \leq \SI{0.19}{\h\per\mpc}$ for standard reconstruction. Moreover, we have $|T(k) - 1| \leq 0.05$ for $k \leq \SI{0.5}{\h\per\mpc}$, demonstrating that our reconstructed initial overdensity is well-normalized in addition to being highly correlated with the true initial overdensity.
This result highlights the power of CNNs for learning local transformations -- once the bulk flows sourced over large scales are corrected by a simple reconstruction algorithm, our CNN produces dramatically improved reconstruction at all scales.

\subsection{Redshift space reconstruction}\label{sec:results-redshift}

Next we considered the problem of reconstructing the \textit{real-space} initial overdensity field from the \textit{redshift-space} overdensity field. This task is more challenging because it requires simultaneously correcting redshift distortions and gravitational collapse.

The red dotted lines in Figure~\ref{fig:corrs_redshift} show the correlation coefficient and transfer function when our CNN is trained to reconstruct the initial real-space overdensity field from the late-time redshift-space overdensity field after preprocessing with standard reconstruction. We refer to this as our ``base method.'' Our base method gives a significant improvement over standard reconstruction in terms of both the correlation coefficient and transfer function.
However, these metrics are averaged over all directions and therefore do not probe the anisotropy of the reconstructed field. Redshift distortions introduce anisotropies that should be corrected in a successful reconstruction. Figure~\ref{fig:quadrupole} shows the quadrupole of the power spectrum and 2-point correlation function of the reconstructed fields. Our base method produces a quadrupole much closer to the true initial field than the late-time field at all scales, and closer than the output of standard reconstruction at all but the largest scales.
However, our base method produces a small anisotropy in the redshift direction at large scales ($k \lesssim \SI{0.2}{\h\per\mpc}$, $r \gtrsim \SI{35}{\per\h\mpc}$), observed as a deviation from the quadrupole of the true initial field.
This undesirable feature is unexpected because our target grids are isotropic: the training algorithm should drive the CNN to produce isotropic outputs on average.
Moreover, correcting redshift distortions on large scales, where the evolution is still in the linear regime, should be \textit{easier} than on small scales, where the evolution is in the nonlinear regime.
We extended our base method in two ways to address this large-scale anisotropy, which we will now present in turn.

First, consider the task of reconstructing the initial density of a particular cell. Redshift distortions have displaced the particles that were initially in that cell, but they have \textit{also} distorted the gravitational sources located some distance away. The task of reconstructing the initial positions of the original particles is complicated by these distortions of the gravitational sources (which remain distorted even after applying standard reconstruction).
However, since redshift distortions have zero mean, on large scales the real-space gravitational sources can be recovered by smoothing the redshift-space density field. 
Motivated by this, we extended our base method by providing our CNN with smoothed copies of the late-time overdensity field at various smoothing scales. We smoothed the input overdensity field (after performing standard reconstruction) with isotropic 3D Gaussian functions of width $L$, $2L$, $4L$, $8L$, and $16L$, where $L \approx \SI{3.5}{\per\h\mpc}$ is the width of one cell. We stacked these smoothed grids together with the input overdensity field and passed the resulting $576^3 \times 6$ grid to our CNN.
In principle, we could demand that the CNN learn any such smoothing transformations during training.
However, it is more efficient in terms of training time and model complexity to explicitly provide these transformations.
Our CNN's receptive field extends only 9 cells in each Cartesian direction, so the larger smoothing scales include information beyond what it can access otherwise.
We also tried including other transformations of the input, such as first and second order gradients of the smoothed fields, but we did not see any benefit beyond that of including the smoothed fields. 

Second, we added a regularization term to the loss function to more strongly encourage the optimization algorithm to reconstruct small-$k$ modes. Regularization terms are commonly used to prevent overfitting when training neural networks. For example, $L_2$ regularization adds a term proportional to the squared $L_2$-norm of the parameter vector, which prevents the optimization algorithm from fitting noise in the training set by penalizing large parameter values. Our concern is that the mean squared error loss function emphasizes large-$k$ modes (small separations) at the expense of small-$k$ modes (large separations).
To see this, note that the sum in equation~\eqref{eq:loss-fourier} contains disproportionately many large-$k$ modes compared to small-$k$ modes (the number of discrete modes in a bin of width $dk$ grows as $k^3$). Accordingly, we added a regularization term of the following form to the loss function:
\begin{equation}\label{eq:regularizer-fourier}
    L_\text{reg} =  \frac{1}{N_\text{sims}} \sum_i \frac{1}{V^2} \sum_{\vk} \ft{w}(\vk)^2 \left| \left( \ft{\delta}^{(i)}_\text{recon}(\vk) - \ft{\delta}^{(i)}_\text{init}(\vk) \right) \right|^2,
\end{equation}
where $\ft{w}(\vk)$ is a real even function monotonically decreasing with $k$. By comparing equations~\eqref{eq:loss-fourier}  and~\eqref{eq:regularizer-fourier}, note that $L_\text{reg}$ is like the mean squared error, but with the contributions of wavevector $\vk$ weighted by $\ft{w}(\vk)^2$. In configuration space, this regularization term is 
\begin{equation}\label{eq:regularizer-conv}
    L_\text{reg} = \frac{1}{N_\text{sims}} \sum_i \frac{1}{N_\text{cells}} \sum_{\vx} \left[ w(\vx) \otimes \left( \delta_\text{recon}^{(i)}(\vx) - \delta_\text{init}^{(i)}(\vx) \right) \right]^2,
\end{equation}
where $w(\vx)$ is the inverse Fourier transform of $\ft{w}(\vk)$ and $\otimes$ denotes convolution. We chose $w(\vx)$ to be an isotropic Gaussian with width of 1 cell ($\approx \SI{3.5}{\per\h\mpc}$), meaning that $\ft{w}(\vk)$ is a Gaussian with width $\sim \SI{0.86}{\h\per\mpc}$. When we included the regularization term, the loss function we used was
\begin{equation}
    L = (1 - \lambda) L_\text{MSE} + \lambda L_\text{reg}.
\end{equation}
We set $\lambda = 0.8$, which provided the best balance of flattening the small-$k$ quadrupole terms without reducing the large-$k$ performance in terms of the correlation coefficient and transfer function.

The purple solid lines in Figures~\ref{fig:corrs_redshift} and~\ref{fig:quadrupole} show the performance of our CNN for redshift-space reconstruction with these two changes. The correlation coefficient and transfer function are very similar to before, but now the quadrupole terms in the 2-point correlation function and power spectrum closely match the true initial overdensity.

Our final method for redshift-space reconstruction significantly improves upon standard reconstruction alone. 
Our method produces a reconstructed field that is highly correlated with the true initial field, with $C(k) > 0.95$ for $k \leq \SI{0.35}{\h\per\mpc}$ versus $k \leq \SI{0.16}{\h\per\mpc}$ for standard reconstruction.
Our reconstructed overdensity is well-normalized, with $|T(k) - 1| \leq 0.05$ for $k \leq \SI{0.35}{\h\per\mpc}$.
While standard reconstruction partially corrects redshift distortions, our method almost completely eliminates the quadrupole discrepancy in the power spectrum and 2-point correlation function, returning a near-isotropic reconstructed field.
Our results demonstrate that CNNs can learn to correct redshift distortions in addition to reversing gravitational collapse, making them a compelling option for future applications of reconstruction to real observational data.

\subsection{Changing the cosmology}

\def\subfigwidth2{0.33\textwidth}
\begin{figure*}
    \centering
    \includegraphics[width=\subfigwidth2]{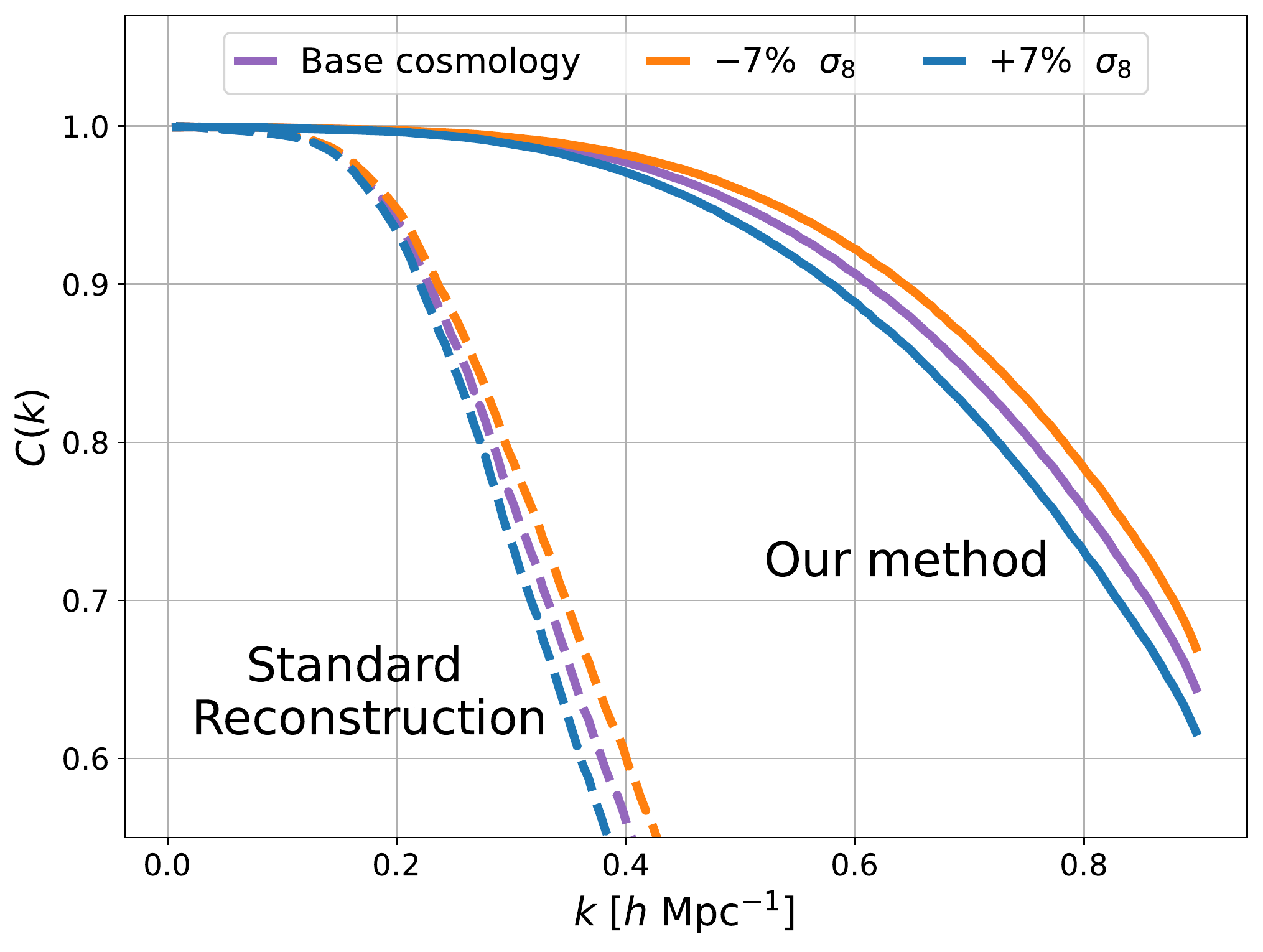}
    \hfill
    \includegraphics[width=\subfigwidth2]{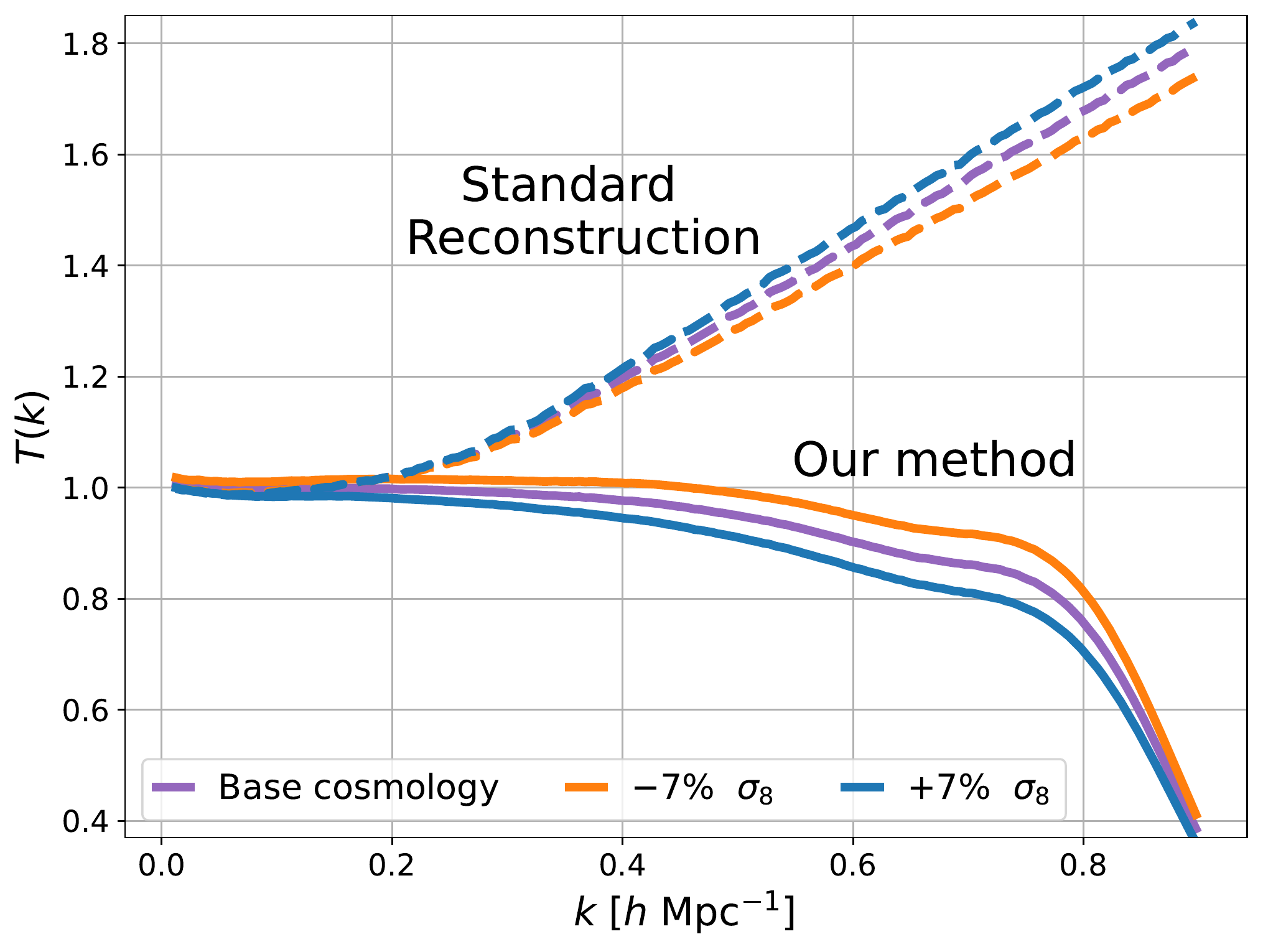}
    \hfill
    \includegraphics[width=\subfigwidth2]{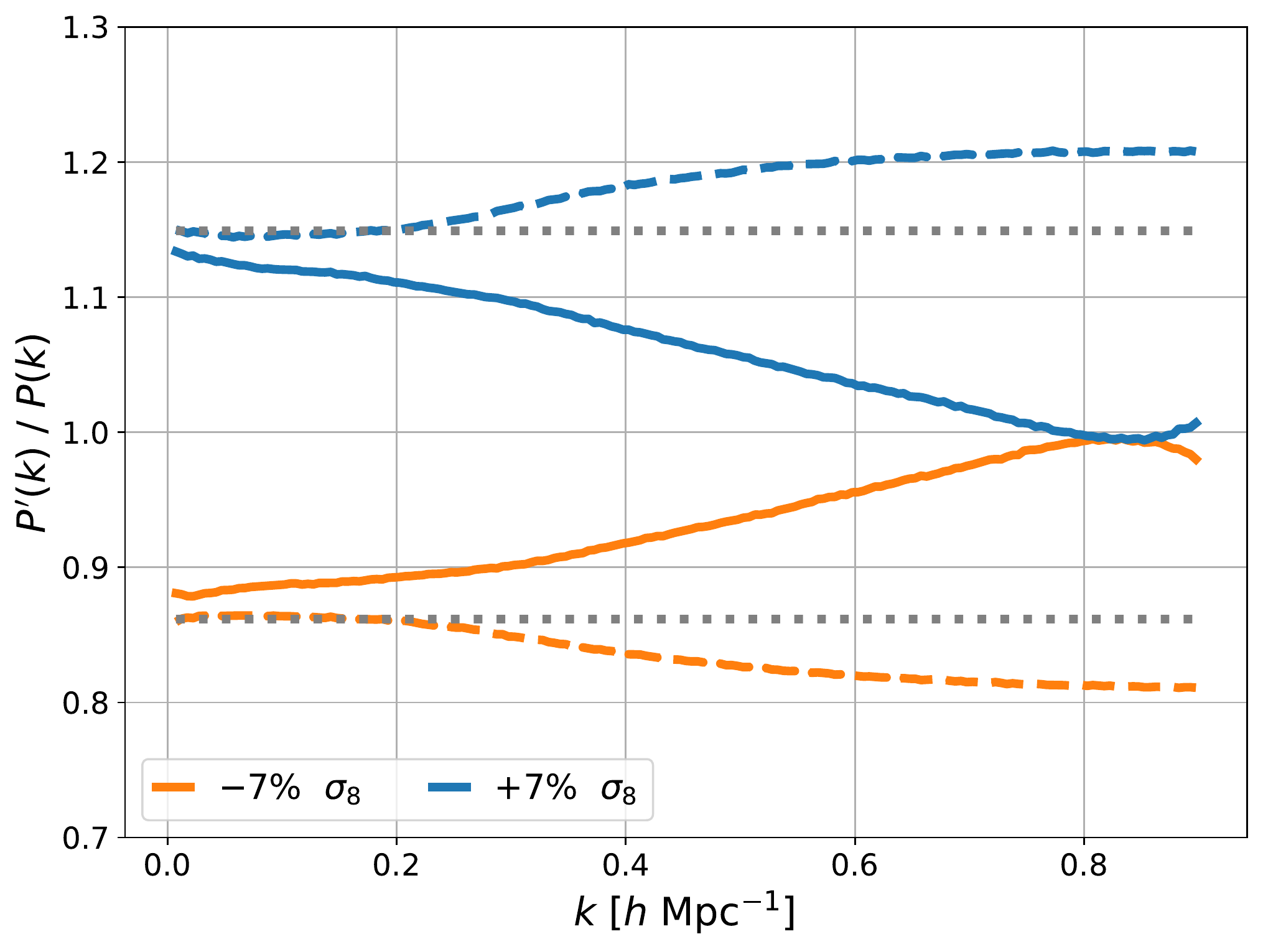}
    \includegraphics[width=\subfigwidth2]{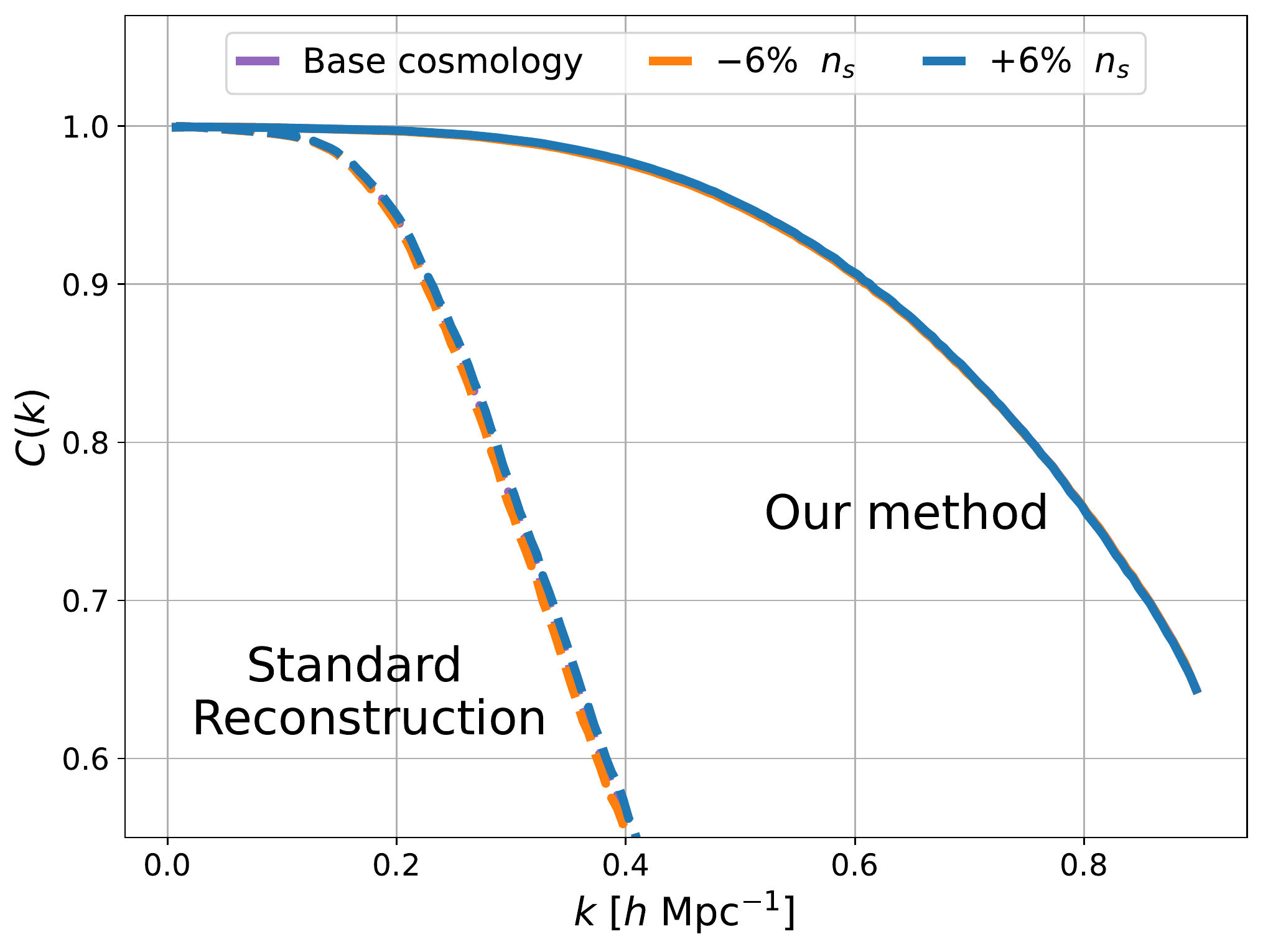}
    \hfill
    \includegraphics[width=\subfigwidth2]{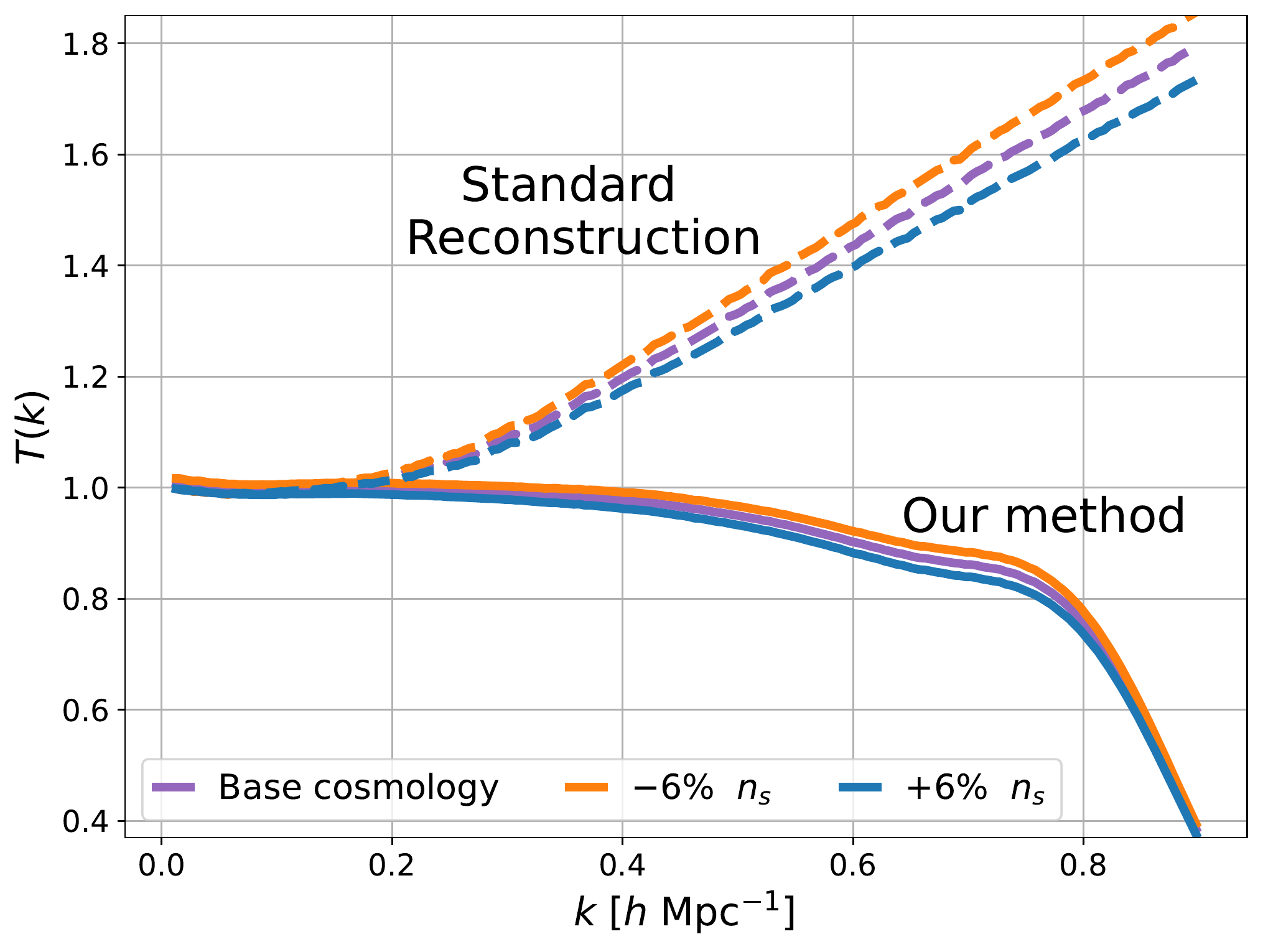}
    \hfill
    \includegraphics[width=\subfigwidth2]{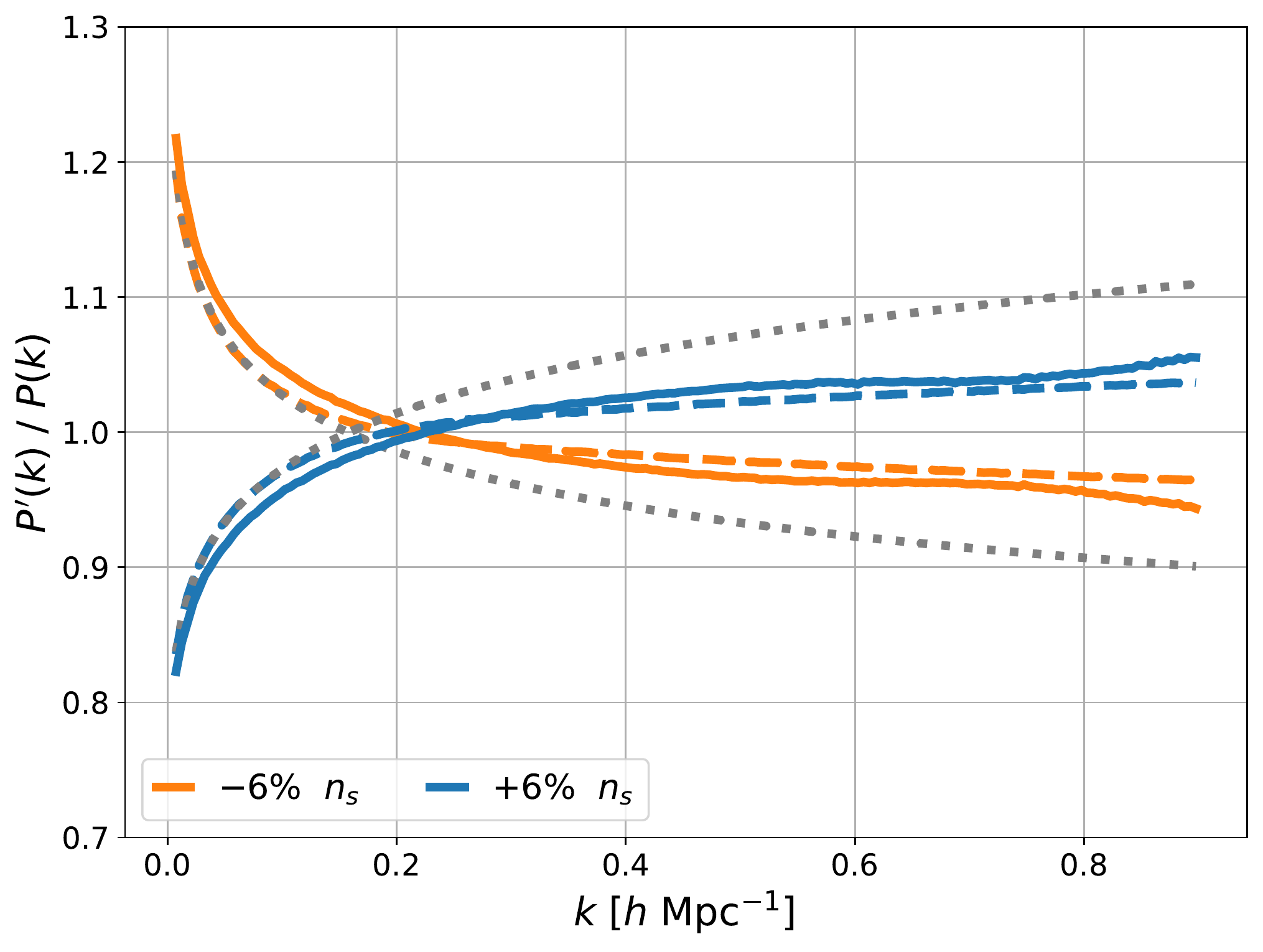}

    \caption{\red Left column: correlation coefficient with respect to the true initial overdensity field ($z=99)$ for simulations where $\sigma_8$ or $n_s$ differs from the training value. Dashed lines are for standard reconstruction, solid lines are for our method.
    Middle column: same as the left column, but for the transfer function.
    Right column: ratio of the reconstructed power spectra of the perturbed and training cosmologies. Dashed lines are for standard reconstruction, solid lines are for our method, dotted lines indicate the true ratio of the initial power spectra.
    }\label{fig:other-cosmologies}
\end{figure*}

The true cosmology of our Universe will inevitably differ from the cosmology used to train our CNN. Accordingly, we must understand how the reconstruction changes when cosmological parameters of the input data differ from those of the training data.
{\red
In this section, we will consider two cosmological parameters that modify the power spectrum of the initial overdensity field: $\sigma_8$, which determines its amplitude, and the spectral index $n_s$, which determines its shape.
Modifying these parameters will change the statistical properties of both the initial and final overdensity fields.
For example, increasing $\sigma_8$ increases the magnitude of initial density perturbations, resulting in a final state with collapsed regions of higher density. Even though the increase in the initial power spectrum is the same for all wavenumbers, the increase in the final power spectrum will be wavenumber-dependent because gravitational collapse is nonlinear.
}
If we apply a CNN trained with a particular value of $\sigma_8$ to a cosmology with a higher value, both the inputs and desired outputs will have different properties compared to those in the training set, and moreover the relationship between inputs and outputs will have changed nonlinearly.
This constitutes a \textit{domain shift}: we cannot expect a machine learning model to perform well on a task that differs considerably from its training task, but we can hope that it still performs well for small changes around the training distribution.

{\red
To evaluate the effects of varying these parameters, we used our CNN trained on the base cosmology from Section~\ref{sec:results} to reconstruct the initial overdensity field for simulations with $\sigma_8$ perturbed by $\pm 7\%$ and simulations with $n_s$ perturbed by $\pm 6\%$.
For clarity, we consider real-space reconstruction to avoid the complicating effects of redshift distortions on the power spectrum and of cosmological parameters and reconstruction on the redshift distortions. The key takeaways are the same in both real space and redshift space.
}

{\red
The left column of Figure~\ref{fig:other-cosmologies} shows the correlation coefficient of the reconstructed initial overdensity with respect to the true initial overdensity for the different values of $\sigma_8$ and $n_s$.
For our CNN, the correlation coefficient decreases with respect to the training cosmology for the larger value of $\sigma_8$: gravitational collapse is further into the nonlinear regime, making reconstruction more difficult.
On the other hand, the correlation coefficient \textit{increases} for the smaller value of $\sigma_8$: the CNN achieves higher fidelity reconstruction even though it was trained with a different value of $\sigma_8$.
The correlation coefficient does not change appreciably when using our CNN on simulations with values of $n_s$ that differ by $\pm 6\%$ from the training cosmology.
}

{\red
The middle column of Figure~\ref{fig:other-cosmologies} shows the transfer function when varying $\sigma_8$ and $n_s$ with respect to the training cosmology.
For our CNN, the transfer function decreases with respect to the training cosmology for the larger value of $\sigma_8$ and vice-versa for the smaller value. The CNN appears to have learned a prior on normalization from the training cosmology, resulting in more underestimated magnitudes when $\sigma_8$ is larger than the base cosmology and more overestimated magnitudes when it is smaller.
We observe a similar effect when changing $n_s$: for wavenumbers where the perturbed power spectrum is greater than the training cosmology ($k \gtrsim \SI{0.2}{\h\per\mpc}$), the CNN's transfer function is lower than that of the training cosmology, and vice-versa.
}

{\red
To dig further into the change in reconstructed power with varying $\sigma_8$ and $n_s$, the right column of Figure~\ref{fig:other-cosmologies} shows the ratio  ${P'_\text{recon}(k)}/{P_\text{recon}(k)}$,
where $P'_\text{recon}(k)$ and $P_\text{recon}(k)$ denote the power spectra of the reconstructed overdensity fields in the perturbed and training cosmologies, respectively.
For small $k$, the magnitude of our CNN's reconstructed power spectrum changes by about the same factor as the true change in the initial power. For larger $k$, its power spectrum becomes increasingly closer to the training cosmology than the perturbed cosmology, indicating a learned prior on power that becomes stronger at smaller scales.
By comparison, standard reconstruction appears relatively more robust than our CNN to changes in $\sigma_8$ and responds similarly to our CNN when varying $n_s$. 
}

{\red
Ultimately, as the left two columns of Figure~\ref{fig:other-cosmologies} show, our CNN significantly improves upon standard reconstruction even if the true cosmology differs from the training cosmology.
However, we consider its tendency to favor the power of the training cosmology as a potential limitation that must be monitored and hopefully mitigated as our method is developed for more real-world applications.
}

\section{Conclusions}\label{sec:conclusions}

Reconstruction aims to give us a window back in time, to an era before gravity molded the Universe we observe today. Revealing the state of matter in the early Universe can allow us to probe effects that have been distorted or obscured by gravitational collapse, such as investigating dark energy by measuring the BAO signature to high precision. A high-fidelity view of the initial state may even yield unanticipated discoveries.

Traditional approaches to reconstruction typically rely on explicit forward or inverse modeling of gravitational dynamics. Machine learning offers an alternative approach: use large-scale cosmological simulations to learn how to transform final conditions directly into initial conditions.
Convolutional neural networks are highly efficient at transforming high-dimensional, structured inputs by assuming that the transformation is local.
However, CNNs are not well-suited for reconstructing the initial density \textit{directly} from the final density because gravity is a long-range force: the relationship between initial and final density is not local.
In this paper we proposed a new method that applies standard reconstruction as a preprocessing step and then uses a CNN to map the partially-reconstructed overdensity field to the true initial overdensity field. The preprocessing step reverses large-scale bulk gravitational flows, transforming the residual reconstruction task from long-range to local and making it well-suited for treatment with a CNN.

We demonstrated that our method performs significantly better than standard reconstruction alone when given either the real-space or redshift-space final overdensity field as input.
For redshift space reconstruction, we extended our base method in two ways to ensure isotropy of the reconstructed field: we augmented the input grid with copies smoothed at different scales to improve the CNN's view of gravitational sources, and we added a regularization term to the loss function to more strongly encourage the optimization algorithm to reconstruct small-$k$ modes.
In both real space and redshift space, our method improves the range of scales of high-fidelity reconstruction ($C(k) > 0.95$) by a factor of 2, corresponding to a factor of 8 increase in the number of well-reconstructed modes. In redshift space, our method almost completely eliminates the quadrupole discrepancy in the power spectrum and 2-point correlation function, returning a near-isotropic reconstructed field.

Future work is needed to account for additional observational limitations that would be present in a galaxy survey.
A surveyed density field will be non-rectangular and, depending on the regions of the sky selected for observation, possibly non-contiguous. Moreover, bright foreground objects, mechanical limitations, and instrumental failures may prevent full coverage of the selected regions. These effects will yield a density field with complicated internal and external boundaries.
Away from such boundaries, no change to the CNN is needed because it only considers input points within its receptive field.
A simple solution would be to assume the mean cosmic density in unobserved regions \citep[as done by][]{mao_baryon_2020}, but better performance near boundaries might be achieved by providing the CNN with a mask indicating which regions of the input were observed and which were unobserved.
Accurate reconstruction near survey boundaries would give CNNs another advantage over traditional methods, since adverse boundary effects can extend up to \SI{100}{\per\h\mpc} for algorithms based on perturbation theory, such as standard reconstruction \citep{zhu_reconstruction_2020}.

{\red
Other key challenges will arise from using galaxies as tracers of the matter field.
Galaxies are much sparser than the dark matter particles used in this paper, so using galaxies will introduce significantly more Poisson noise in the density fields.
Moreover, the galaxy density is a biased estimator of the true matter density; the difference between these density fields is characterized by a set of \textit{galaxy bias parameters} \citep{desjacques_large-scale_2018}.
In principle, the bias parameters could be estimated independently and the galaxy density field corrected before applying reconstruction. However, we could also attempt to correct galaxy bias within our CNN framework.
Assuming we do not know the bias parameters of the Universe \textit{a priori}, we would need to train the CNN to detect the bias itself.
This would necessitate extending its architecture in some way -- currently, our CNN only observes local patches of the input field, but detecting the bias parameters would likely require global information, for example summary statistics that could be fed in as separate inputs.
With an architecture capable of detecting and correcting galaxy bias, we could then train the CNN on simulated galaxy fields generated from many different possible values of the bias parameters.
We leave these extensions to future work.
Since the magnitude of galaxy bias increases with redshift, accounting for galaxy bias will become even more important for future galaxy surveys that will focus on $z>1$.
}

As current and future galaxy surveys map the local Universe in greater detail than ever before, new computational techniques will be needed to get the most out of this data. CNNs promise to be a powerful tool in this toolkit, presenting opportunities to untangle the non-linear clustering at intermediate scales more accurately than traditional methods.

\section*{Acknowledgements}

CJS and DJE were supported by U.S. Department of Energy grant DE-SC0013718 and by the National Science Foundation under Cooperative Agreement PHY-2019786 (the NSF AI Institute for Artificial Intelligence and Fundamental Interactions, \url{http://iaifi.org/}).  DJE is further supported as a Simons Foundation Investigator.


\section*{Data Availability}

The \abacussummit\ suite of cosmological simulations, the source of all training and evaluation data in this paper, is available for download at \url{https://abacusnbody.org}. The persistent DOI for the \abacussummit\ data release is 10.13139/OLCF/1811689.
The code used to generate the results in this paper is available at \url{https://github.com/cshallue/recon-cnn}. The state of the code at submission time is archived at \url{https://doi.org/10.5281/zenodo.6856562}.
The trained CNN models and the data underlying the plots in this paper are available at \url{https://doi.org/10.5281/zenodo.7498063}.


\bibliographystyle{mnras}
\bibliography{ReconCNN}






\bsp	
\label{lastpage}
\end{document}